\documentclass[%
 reprint,
 amsmath,amssymb,
aps,
pre,
nofootinbib]{revtex4-2}

\usepackage{graphicx}% Include figure files
\usepackage{dcolumn}% Align table columns on decimal point
\usepackage{bm}% bold math
\usepackage{amsmath}	% Advanced maths commands
\usepackage{amssymb}	% Extra maths symbols
\usepackage{txfonts}
\usepackage[bb=libus]{mathalpha}
\usepackage{bbold}
\usepackage{relsize}
\usepackage{mathtools}
\usepackage{bigints}
\usepackage{sidecap}
\usepackage[caption=false]{subfig}
\usepackage{hyperref}

\usepackage{floatrow}
\usepackage{soul}
\usepackage{comment}
\usepackage[normalem]{ulem}

\usepackage{macros_ub}
\graphicspath{{./}{figures/}}

\begin{document}

\preprint{APS/123-QED}

\title{Relaxation of weakly collisional plasma: continuous spectra, discrete eigenmodes, and the decay of echoes}% Force line breaks with \\
%\thanks{A footnote to the article title}%

\author{Uddipan Banik}
\email{uddipan.banik@princeton.edu,\;uddipanbanik@ias.edu}
\affiliation{Department of Astrophysical Sciences, Princeton University, 112 Nassau Street, Princeton, NJ 08544, USA\\School of Natural Sciences, Institute for Advanced Study, Einstein Drive, Princeton, NJ 08540, USA\\Perimeter Institute for Theoretical Physics, 31 Caroline Street N., Waterloo, Ontario, N2L 2Y5, Canada}
 %\altaffiliation[Also at ]{}%Lines break automatically or can be forced with \\
\author{Amitava Bhattacharjee}%
\email{amitava@princeton.edu}
\affiliation{Department of Astrophysical Sciences, Princeton University, 112 Nassau Street, Princeton, NJ 08540, USA\\
%
 %Authors' institution and/or address\\
 %This line break forced with \textbackslash\textbackslash
}%

%\collaboration{MUSO Collaboration}%\noaffiliation

%\author{Charlie Author}
% \homepage{http://www.Second.institution.edu/~Charlie.Author}
%\affiliation{
% Second institution and/or address\\
% This line break forced% with \\
%}%
%\affiliation{
% Third institution, the second for Charlie Author
%}%
%\author{Delta Author}
%\affiliation{%
% Authors' institution and/or address\\
% This line break forced with %\textbackslash\textbackslash
%}%

%\collaboration{CLEO Collaboration}%\noaffiliation

\date{\today}% It is always \today, today,
             %  but any date may be explicitly specified

\begin{abstract}

The relaxation of a weakly collisional plasma, which is of fundamental interest to laboratory and astrophysical plasmas, can be described by the self-consistent Boltzmann-Poisson equations with the Lenard-Bernstein collision operator. We perform a perturbative (linear and second order) analysis of the Boltzmann-Poisson equations, and obtain exact analytic solutions which resolve some long-standing controversies regarding the impact of weak collisions on the continuous spectra, the discrete Landau eigenmodes, and the decay of plasma echoes. We retain both damping and diffusion terms in the collision operator throughout our treatment. We find that the linear response is a temporal convolution of two types of contribution: a continuum that depends on the continuous velocities of particles (crucial for the plasma echo), and another, consisting of discrete modes that are coherent modes of oscillation of the entire system. The discrete modes are exponentially damped over time due to collective effects or wave-particle interactions (Landau damping), as well as collisional dissipation. The continuum is also damped by collisions, but somewhat differently than the discrete modes. Up to a collision time, which is the inverse of the collision frequency $\nu_\rmc$, the continuum decay is driven by the diffusion of particle velocities and is cubic exponential, occurring over a timescale $\sim \nu^{-1/3}_\rmc$. After a collision time, however, the continuum decay is driven by the collisional damping of particle velocities and diffusion of their positions, and occurs exponentially over a timescale $\sim \nu_\rmc$. This slow exponential decay causes perturbations to damp the most on scales comparable to the mean free path, but very slowly on larger scales. This establishes the local thermal equilibrium (LTE), which is the essence of the fluid limit. The long-term decay of the linear response is driven by the discrete modes on scales smaller than the mean free path, but, on larger scales, is governed by a combination of the slowly decaying continuum and the least damped discrete mode. This slow exponential decay implies that the echo, which results from the interference of the continuum response to two subsequent pulses, is detectable even on scales comparable to the mean free path, as long as the second pulse is introduced within a few phase-mixing timescales after the first.

%\begin{description}
%\item[Usage]
%Secondary publications and information retrieval purposes.
%\item[Structure]
%You may use the \texttt{description} environment to structure your abstract;
%use the optional argument of the \verb+\item+ command to give the category of each item. 
%\end{description}
\end{abstract}

%\keywords{Suggested keywords}%Use showkeys class option if keyword
                              %display desired
\maketitle

%\tableofcontents

\section{Introduction}

The relaxation of plasma is a complicated many-body problem with rich phenomenology. Many astrophysical and space plasmas are nearly collisionless and are often described by the collisionless Boltzmann (also known as the Vlasov) equation, widely discussed in textbooks. However, no matter how infrequent the collisions, no plasma in nature or a terrestrial laboratory is strictly collisionless, so in practical terms one may also treat a collisionless plasma as the zero-collision limit of a weakly collisional plasma. Furthermore, if the collision operator is of the Fokker-Planck type, the collision frequency typically multiplies the highest velocity derivative in the kinetic equation. Taking the limit of zero collisions is then a subtle problem, requiring the use of singular perturbation theory. It should then not be surprising if the results obtained in the zero-collision limit of the weakly collisional problem differ qualitatively from those obtained by solving the collisionless problem. The physical effect of weak collisions is to regulate extreme filamentation in phase space, much in the same way as viscosity tends to regulate near-singular vortex structures in fluids governed by the Navier-Stokes equations.

These considerations have deep implications for the thermodynamics of many-body systems. It is well known that the Vlasov equation does not possess an H-theorem, but the Boltzmann equation does. The Vlasov equation has a denumerably infinite number of Casimirs (invariants). How many of them (if any) survive in the weakly collisional limit? And how do they influence the turbulent relaxation of a weakly collisional plasma? The Vlasov equation, despite the occurrence of Landau damping, is time-reversible, as demonstrated by the phenomenon of plasma echoes. How do weak collisions affect the time-evolution of echoes? 

The questions posed above do not just belong to the esoteric realm of mathematical theory but are germane to a variety of weakly collisional astrophysical and space plasma phenomena, which can be described using kinetic theory. For example, the problem of decay of plasma echoes in the presence of weak collisions, which is a fundamental problem in plasma physics, has important implications for the role echoes can play in competing with phase-mixing and Landau damping in plasma turbulence \citep[][]{Schekochihin.etal.16}. This, in turn, has significant implications for understanding the energetics of plasma turbulence and interpreting observations and numerical experiments pertaining to, for example, solar wind heating and acceleration \citep[][]{Bruno.Carbone.05, Marsch.06, Howes.17}, reconnection in a nearly collisionless magnetosphere \citep[][]{Burch.etal.16}, formation of coherent structures such as electron holes in phase space \citep[][]{Hutchinson.17}, or particle heating near a black hole \citep[][]{Chael.etal.18}. Of course, most of these scenarios involve large perturbations and non-linear effects, for which numerical simulations and theories of turbulent cascade \cite{Nastac.etal.23} are the way forward. However, before addressing the fully nonlinear problem, we must have a holistic understanding of the perturbative relaxation of weakly collisional plasmas, especially the spatio-temporal evolution of the distribution function, coupled self-consistently with the Poisson equation (if the plasma dynamics is assumed to be electrostatic). This paper addresses some of the open issues on this topic.

Understanding the modal response of a weakly collisional plasma is critical to the development of a statistical description of the system (be it in equilibrium or non-equilibrium). Recently, there has been a resurgence of interest in the long-time  behavior of weakly collisional kinetic plasmas. A rigorous linear theory of weakly collisional plasmas \citep [][] {Ng.Bhattacharjee.Skiff.99,Ng.Bhattacharjee.Skiff.04,Ng.Bhattacharjee.21}, which builds on the seminal work of \cite{Lenard.Bernstein.58}, demonstrated that a plasma governed by the Lenard-Bernstein (LB) collision operator (a special case of the Fokker-Planck operator) oscillates in the singular limit of zero collisions as a complete set of discrete normal modes at long times. A striking finding of these studies was that the Case-van Kampen modes, which are known to be a complete set of normal modes for a purely collisionless plasma and constitute a continuous spectrum, are eliminated when weak collisions are introduced into the system. The new eigenmodes of a weakly collisional plasma constitute a discrete spectrum, and include the Landau solutions. As is well known, the Landau solutions are not eigenmodes of the collisionless Vlasov-Poisson system and are instead remnants of the initial perturbations in the time-asymptotic limit. However, in the presence of weak collisions, these become eigenmodes (referred to hereafter as Landau modes\footnote{In the collisionless limit, on scales smaller than the Debye length, these are Landau damped, while, on scales larger than the Debye length, these are nothing but the weakly damped electron Langmuir waves and ion acoustic waves.}). For a stable plasma, these eigenmodes undergo a simple exponential decay with time. On the other hand, \cite{Su.Oberman.68} showed in a classic paper that the linear response to perturbations decays cubic exponentially due to collisions (at least at short time). Although they do not make it explicit in their paper, their analysis only applies to the continuum part of the plasma response, as we clarify in section~\ref{sec:cont_resp}. Their findings imply that the plasma echo, which arises non-linearly from the continuum response, would rapidly decay due to electron-ion collisions. The \cite{Su.Oberman.68} prediction for the temporal decay of the continuum response is, however, much more rapid than the exponential decay of the discrete eigenmodes, identified by \cite{Ng.Bhattacharjee.Skiff.99,Ng.Bhattacharjee.Skiff.04}. \cite{Short.Simon.02} revisited the \cite{Su.Oberman.68} result from a boundary-layer analysis of the problem, reaffirming the results of \cite{Su.Oberman.68}.

It thus appears that the relaxation of weakly collisional plasmas has produced somewhat conflicting results: the super-exponential (cubic exponential) decay of the continuum response and the plasma echo according to \cite{Su.Oberman.68} and the exponential decay of the discrete eigenmodes according to \citep[][]{Lenard.Bernstein.58,Ng.Bhattacharjee.Skiff.99,Ng.Bhattacharjee.Skiff.04,Black.etal.13,Ng.Bhattacharjee.21}. This naturally raises the following questions. Which of these results is correct? And, if both are correct, are these results simply different limits of the same response? This paper addresses these very questions. We show that the exact solution to the problem of complete perturbative (linear and second order) relaxation of a weakly collisional plasma indeed yields the above results as different asymptotic limits in time. We show that including the damping (also known as dynamical friction/drag \citep[][]{Chandrasekhar.43}) term in the collision operator modifies the \cite{Su.Oberman.68} prediction of cubic exponential decay of the continuum response to an exponential decay at late times and large spatial scales. We find that the \cite{Su.Oberman.68} prediction only holds for short times, i.e., $t<1/\nu_\rmc$, and scales smaller than the mean free path, i.e., $\lambda<\lambda_\rmc = \sigma/\nu_\rmc$ or $\nu_\rmc < k\sigma$, where $\nu_\rmc$ is the collision frequency, $\sigma$ is the equilibrium velocity dispersion, $\lambda_\rmc = \sigma/\nu_\rmc$ is the mean free path, $k$ is the wavenumber of the response and $\lambda = 2\pi/k$ is the wavelength. At long times ($t>1/\nu_\rmc$) and on large scales ($\lambda>\lambda_\rmc$ or $\nu_\rmc>k\sigma$), the response damps exponentially and much slower, at a rate similar to that predicted by \citep[][]{Ng.Bhattacharjee.Skiff.99,Ng.Bhattacharjee.Skiff.04}. The damping is most pronounced for $\nu_\rmc \sim k\sigma$ or $\lambda \sim \lambda_\rmc$. The slow exponential decay outside the mean free path implies that the distribution function (DF) approaches the equilibrium Maxwellian form rapidly only within the mean free path; this is consistent with the assumption of the establishment of a local thermodynamic equilibrium (LTE) in the fluid approximation (large $\nu_\rmc$). Hence, we recover the fluid limit from a rigorous theory of weakly collisional plasmas.

We remark that the presence of two timescales in a problem of kinetic relaxation is, by itself, not a novelty. For example, as pointed out to us by an anonymous referee, the problem has been extensively investigated in the context of inertial confinement fusion (ICF) experiments which require drastic revisions of classical transport theory, and also provide two temporal behaviors, one which is super-exponential and the other, exponential (see \citep[][]{Bell.83,Epperlein.etal.92,Bychenkov.etal.95,Brantov.Bychenkov.06,Brantov.Bychenkov.12,Rozmus.etal.17} and other references therein). However, the time-dependence as well as the parametric dependence of the kinetic (collisionless) super-exponential decay of the linear response on plasma parameters obtained in these works are quite different from our findings on the weakly collisional regime of the linear response and the non-linear plasma echoes. For example, the kinetic super-exponential decay of the density and temperature perturbations found by \citep[]{Brantov.Bychenkov.06,Rozmus.etal.17} is Gaussian, whereas we find a more general form of the decay in the weakly collisional regime that can be Gaussian in one limit and cubic exponential (same as \citep[][]{Su.Oberman.68}) in another (see section~\ref{sec:combined_lin_resp} for a detailed discussion on this). The slow exponential decay of the linear response and the non-linear echoes in the fluid regime is, on the other hand, in good agreement with the slow decay of the linear response in this regime obtained by \citep[][]{Bell.83,Epperlein.etal.92,Bychenkov.etal.95,Brantov.Bychenkov.06,Brantov.Bychenkov.12,Rozmus.etal.17}. The classical collisional theory of plasma echoes \citep[][]{Su.Oberman.68} has suffered from the absence of this slow exponential decay, a defect we remedy in this paper. Furthermore, a key aspect of our analysis that differs from \citep[]{Bell.83,Epperlein.etal.92,Bychenkov.etal.95,Brantov.Bychenkov.06,Brantov.Bychenkov.12,Rozmus.etal.17} is that we go beyond linear theory, and perform a second-order treatment of the kinetic equations to characterize the collisional damping of the plasma echo.

We organize the paper as follows. In section~\ref{sec:lin_resp_theory}, we discuss the perturbative response theory of a weakly collisional plasma. In section~\ref{sec:lin_eq_solve}, we briefly sketch the solution of the linearized Boltzmann-Poisson equations assuming an LB collision operator. Section~\ref{sec:cont_resp} describes the detailed behaviour of the continuum response along with a complementary analysis using the Langevin formalism. In section~\ref{sec:normal_modes}, we describe the discrete mode response, and discuss the damping rates and oscillation frequencies of the various modes. In section~\ref{sec:combined_lin_resp}, we compute the complete linear response of a weakly collisional plasma, which is a temporal convolution of the continuum response and the discrete mode response. In Section~\ref{sec:echo}, using second-order response theory, we discuss the impact of weak collisions on the observability of the plasma echo in experiments, which is shown to be a manifestation of the continuum response. We summarize our findings and its possible implications for a variety of applications in section~\ref{sec:discussion_summary}.

\section{Perturbative response theory for weakly collisional plasma}\label{sec:lin_resp_theory}

A plasma is characterized by the DF or phase space ($\bx,\bv$) density of charged particles, $f(\bx,\bv,t)$. Typically, the electrons are much less massive and thus more mobile than the ions, and are the dominant drivers of plasma oscillations. Hence, we shall restrict ourselves to the study of the evolution of the electron DF, with the ions forming a stationary, homogeneous and cold background. The general equations governing the evolution of a (non-magnetized) electrostatic plasma are the Boltzmann-Poisson equations. The Boltzmann equation,

\begin{align}
\frac{\partial f}{\partial t} + \bv\cdot{\bf \nabla} f + \frac{e}{m}{\bf \nabla}_\bv f \cdot {\bf \nabla}\left(\Phi+\Phi_\rmP\right) = C\left[f\right],
\end{align}
is a conservation equation for the DF. Here, $e$ is the magnitude of electric charge and $m$ is the electron mass. $C[f]$ denotes the collision operator, $\Phi_\rmP$ is the electrostatic potential of an external perturbation, and $\Phi$ is the self-potential of the plasma, which is sourced by the electron DF via the Poisson equation:

\begin{align}
\nabla^2\Phi = \frac{e}{\epsilon_0} \left(\int d^3v\, f - n_e\right),
\end{align}
with $n_e$ the equilibrium electron (or ion) density.

If the strength of the perturber potential, $\Phi_\rmP$, is smaller than $\sigma^2$, where $\sigma$ is the velocity dispersion of the unperturbed equilibrium plasma (typically characterized by a Maxwellian DF), then the perturbation in $f$ can be expanded as a power series in the perturbation parameter, $\varepsilon \sim \left|\Phi_\rmP\right|/\sigma^2$, i.e., $f = f_0 + \varepsilon f_1 + \varepsilon^2 f_2 + ...\,$. The evolution equations for the linear order perturbation in the DF, $f_1$ (which we shall henceforth refer to as the linear response), and that in the plasma potential, $\Phi_1$, are given by the following linearized form of the Boltzmann-Poisson equations:

\begin{align}
&\frac{\partial f_1}{\partial t} + \bv\cdot{\bf \nabla}f_1 = -\frac{e}{m}{\bf \nabla}_{\bv} f_0 \cdot{\bf \nabla}\left(\Phi_\rmP+\Phi_1\right) + C\left[f_1\right],\nonumber\\
&\nabla^2 \Phi_1 = \frac{e}{\epsilon_0} \int d^3v f_1.
\label{lin_BE_Poisson}
\end{align}
Similarly, the evolution equations for the second order perturbations, $f_2$ and $\Phi_2$, are given by

\begin{align}
&\frac{\partial f_2}{\partial t} + \bv\cdot{\bf \nabla}f_2 = -\frac{e}{m}{\bf \nabla}_{\bv} f_1 \cdot{\bf \nabla}\left(\Phi_\rmP+\Phi_1\right) -\frac{e}{m}{\bf \nabla}_{\bv} f_0 \cdot{\bf \nabla}\Phi_2\nonumber\\
& \hspace{1.9cm} + C\left[f_2\right],\nonumber\\
&\nabla^2 \Phi_2 = \frac{e}{\epsilon_0} \int d^3v f_2.
\label{2nd_BE_Poisson}
\end{align}
We discuss the evolution of the linear response in sections~\ref{sec:lin_eq_solve}, \ref{sec:cont_resp}, \ref{sec:normal_modes} and \ref{sec:combined_lin_resp}. We discuss the second order response in section~\ref{sec:echo} in the context of the plasma echo phenomenon.

We assume that the collisional scattering of electrons is predominantly driven by weak encounters with other electrons and ions, i.e., collisions only cause small deflections from the unperturbed collisionless orbits. Under this assumption, the collision operator, $C[f_1]$, can be approximated as the following Fokker-Planck operator:

\begin{align}
C[f_1] \approx \frac{1}{2} \frac{\partial}{\partial \chi_i} \left(D_{ij} \frac{\partial f_1}{\partial \chi_j}\right),
\end{align}
where ${\mathbb {\chi}} = \left({\bx, \bv}\right)$ denotes the canonical phase space coordinates, and $D_{ij}$ denote the diffusion tensor ($i,j=1(1)3$).

\section{Solving the linearized equations}\label{sec:lin_eq_solve}

We solve the linearized problem (equations~[\ref{lin_BE_Poisson}]) under a set of simplifying assumptions. First, we assume that the unperturbed system is homogeneous and characterized by an equilibrium DF, $f_0(v)$. Second, we restrict ourselves to $1$D perturbations. Third, we assume that the majority of collisions are weak, i.e., only cause slight deflections of the electrons from their orbits. This entails that the collision operator can be simplified into the Fokker Planck form. One such operator that gives rise to a Maxwellian DF in equilibrium, given by

\begin{align}
f_0(v) = \frac{n_e}{\sqrt{2\pi}\sigma} \exp{\left[-\frac{v^2}{2\sigma^2}\right]},
\end{align}
is attributed to Lenard and Bernstein (LB) \citep[][]{Lenard.Bernstein.58}, and is given by

\begin{align}
C\left[f\right] = \nu_\rmc \frac{\partial}{\partial v}\left(v f + \sigma^2 \frac{\partial f}{\partial v}\right),
\label{LB_coll_op}
\end{align}
where $\nu_\rmc$ is the collision frequency and $\sigma$ is the velocity dispersion of the system when it is in equilibrium. The Boltzmann equation with the LB operator is also known as the mean-field Kramers equation \citep[see][for a detailed discussion of the initial-value problem for this equation in the context of weakly collisional plasmas and self-gravitating systems]{Chavanis.13}. The fundamental assumption behind this form of the collision operator is that the collisions are sufficiently short-range and random. The LB operator conserves the particle number but not the linear momentum and kinetic energy during collisions. This shortcoming is well-known in the plasma literature \citep[][]{Lenard.Bernstein.58,Ng.Bhattacharjee.Skiff.99,Ng.Bhattacharjee.Skiff.04,Ng.Bhattacharjee.21}. Despite this shortcoming, we expect the scaling behaviours of different aspects of plasma relaxation to be qualitatively robust (see section~\ref{sec:Langevin_analysis} for details).

Under the above assumptions, the linearized Boltzmann equation given by the first of equations~(\ref{lin_BE_Poisson}) becomes the Lenard-Bernstein (LB) equation in $1$D, and can be written as

\begin{align}
\frac{\partial f_1}{\partial t} + v\frac{\partial f_1}{\partial x}  = -\frac{e}{m}\frac{\partial f_0}{\partial v} \frac{\partial}{\partial x}\left(\Phi_1+\Phi_\rmP\right) + \nu_\rmc \frac{\partial}{\partial v}\left(v f_1 + \sigma^2 \frac{\partial f_1}{\partial v}\right).
\label{lin_CBE_act_ang}
\end{align}
To solve this equation simultaneously with the Poisson equation, we take the Fourier transform of both sides in $x$ and the Laplace transform in $t$. We perform a Laplace transform (and not a Fourier transform) in time because we are interested in the initial-value problem. We then obtain

\begin{align}
\left({\gamma+ikv}\right) \Tilde{f}_{1k} &= f_{1k}(0) - ik\left(\Tilde{\Phi}_{1k}+\Tilde{\Phi}_k\right) \frac{e}{m}\frac{\partial f_0}{\partial v} \nonumber\\
&+ \nu_\rmc \frac{\partial}{\partial v}\left(v \Tilde{f}_{1k} + \sigma^2 \frac{\partial \Tilde{f}_{1k}}{\partial v}\right).
\label{master_eq}
\end{align}
Here, $\Tilde{f}_{1k}$, $\Tilde{\Phi}_{1k}$ and $\Tilde{\Phi}_k$ are the Fourier-Laplace transforms of $f_1$, $\Phi_1$ and $\Phi_\rmP$ respectively, i.e., 

\begin{align}
Q(x,t) = \frac{1}{2\pi i} \int_{c-i\infty}^{c+i\infty} \rmd \gamma \int_{-\infty}^\infty \rmd k\, \Tilde{Q}_k(\gamma) \exp{\left[ikx+\gamma t\right]},
\end{align}
where $Q$ refers to each of $f_1$, $\Phi_1$ and $\Phi_\rmP$, while $\Tilde{Q}_k$ refers to each of $\Tilde{f}_{1k}$, $\Tilde{\Phi}_{1k}$ and $\Tilde{\Phi}_k$. Here, $c$ is a real number such that the $\gamma$ integral is performed within its region of convergence. $\Tilde{\Phi}_{1k}$ is related to $\Tilde{f}_{1k}$ through the Poisson equation,

\begin{align}
\Tilde{\Phi}_{1k} = -\frac{e}{\epsilon_0\, k^2} \int \rmd v \,\Tilde{f}_{1k}.
\label{Poisson_eq_Fourier_Laplace}
\end{align}

Let us now non-dimensionalize the LB and Poisson equations (equations~[\ref{master_eq}] and [\ref{Poisson_eq_Fourier_Laplace}]) before proceeding further. Defining $\kappa=k\sigma/\nu_\rmc$, $\eta=\gamma/\nu_\rmc$, $u=v/\sigma$, $\Tilde{g}_{1\kappa} = \Tilde{f}_{1k}(\sigma\nu_\rmc/n_e)$, $g_0=f_0(\sigma/n_e)$, $h_\kappa = f_{1k}(0)(\sigma/n_e)$, $A_\kappa(\eta)=-(\nu_\rmc e/m\sigma^2)\,\Tilde{\Phi}_k(\eta)$, and the plasma frequency, $\omega_\rmP = \sqrt{n_e e^2/m \epsilon_0}$, we have the following non-dimensional form of the linearized LB and Poisson equations:

\begin{align}
\left({\eta+i\kappa u}\right) \Tilde{g}_{1\kappa} &= h_\kappa + i\kappa A_\kappa(\eta)\frac{\partial g_0}{\partial u} + \frac{i}{\kappa}\,{\left(\frac{\omega_\rmP}{\nu_\rmc}\right)}^2 \frac{\partial g_0}{\partial u} \int \rmd u\, \Tilde{g}_{1\kappa} \nonumber\\
&+ \frac{\partial}{\partial u}\left(u \Tilde{g}_{1\kappa} + \frac{\partial \Tilde{g}_{1\kappa}}{\partial u}\right).
\label{master_eq_nd}
\end{align}

The above is an integro-differential equation and thus further simplification is required before a solution is attempted. This crucial simplifying step involves taking the velocity Fourier transform of $\Tilde{g}_{1\kappa}$, $h_\kappa$ and $g_0$ (assuming that these tend to zero sufficiently fast as $u$ tends to $\pm \infty$), i.e., expanding these as

\begin{align}
\Tilde{g}_{1\kappa}(u) &= \int_{-\infty}^\infty \rmd w \, \exp{\left[iwu\right]}\, \Tilde{G}_{1\kappa}(w), \nonumber\\
h_\kappa(u) &= \int_{-\infty}^\infty \rmd w \, \exp{\left[iwu\right]}\, H_\kappa(w), \nonumber\\
g_0(u) &= \int_{-\infty}^\infty \rmd w \, \exp{\left[iwu\right]}\, G_0(w).
\end{align}
This implies that $\int \rmd u \, \Tilde{g}_{1\kappa}(u) = 2\pi \Tilde{G}_{1\kappa}(0)$. This reduces the integro-differential equation~(\ref{master_eq_nd}) to the following first order ordinary differential equation in $w$:

\begin{align}
&(w-\kappa)\,\frac{\partial \Tilde{G}_{1\kappa}}{\partial w} + (w^2+\eta)\,\Tilde{G}_{1\kappa} \nonumber\\
&= H_\kappa(w) - \left[\frac{2\pi}{\kappa}{\left(\frac{\omega_\rmP}{\nu_\rmc}\right)}^2 \Tilde{G}_{1\kappa}(0)+\kappa A_\kappa(\eta)\right]w\, G_0(w).
\label{master_eq_nd_w}
\end{align}
This can be integrated using the method of integrating factor \citep[][]{Lenard.Bernstein.58} to yield the following solution for $\Tilde{G}_{1\kappa}(w)$:

\begin{align}
\Tilde{G}_{1\kappa}(w) &= \exp{\left[-\frac{w^2}{2} - \kappa w\right]}\;\; {\left(w-\kappa\right)}^{-\left(\eta+\kappa^2\right)} \nonumber \\
&\times \int_\kappa^w \rmd w'\, \exp{\left[\frac{w'^2}{2} + \kappa w'\right]}\;\; {\left(w'-\kappa\right)}^{\,\eta+\kappa^2-1} \nonumber\\
&\times \left[H_\kappa(w') - \left(\frac{2\pi}{\kappa}{\left(\frac{\omega_\rmP}{\nu_\rmc}\right)}^2 \Tilde{G}_{1\kappa}(0)+\kappa A_\kappa(\eta)\right)w' G_0(w')\right].
\label{G_1k_w}
\end{align}
Substituting $w=0$ in the above, we get the following expression for $\Tilde{G}_{1\kappa}(0)$:

\begin{align}
\Tilde{G}_{1\kappa}(0) = \frac{N(\kappa,\eta)}{D(\kappa,\eta)}.
\label{G_1k_0}
\end{align}
Here,

\begin{align}
N(\kappa,\eta) &= {\left(-\kappa\right)}^{-\left(\eta+\kappa^2\right)} \int_\kappa^0 \rmd w'\, \exp{\left[\frac{w'^2}{2} + \kappa w'\right]}\nonumber\\
&\times {\left(w'-\kappa\right)}^{\,\eta+\kappa^2-1} \left[H_\kappa(w') -\kappa A_\kappa(\eta) w' G_0(w')\right],\nonumber \\\nonumber \\
D(\kappa,\eta) &= 1 + \frac{2\pi}{\kappa}{\left(\frac{\omega_\rmP}{\nu_\rmc}\right)}^2 {\left(-\kappa\right)}^{-\left(\eta+\kappa^2\right)} \nonumber\\
&\times \int_\kappa^0 \rmd w'\, \exp{\left[\frac{w'^2}{2} + \kappa w'\right]}\;\; {\left(w'-\kappa\right)}^{\,\eta+\kappa^2-1} w' G_0(w') \nonumber\\
&= 1+ {\left(\frac{\omega_\rmP}{\kappa\nu_\rmc}\right)}^2 \left[\,1-\left(a-\zeta\right)\zeta^{-a}\exp{\left[\zeta\right]}\,\gamma\left(a,\zeta\right)\,\right],
\label{N_D_0}
\end{align}
where $a=\eta+\kappa^2$, $\zeta = \kappa^2$, and $\gamma(a,\zeta)=\int_0^\zeta \rmd z\,z^{a-1}\exp{\left[-z\right]}$ is the lower incomplete gamma function. We have assumed that the unperturbed DF, $g_0$, is of Maxwellian form, i.e., $g_0(u) = \exp{\left[-u^2/2\right]}\Big/\sqrt{2\pi}$, whose velocity Fourier transform is given by

\begin{align}
G_0(w) = \frac{1}{2\pi}\int_{-\infty}^\infty \rmd u\, \exp{\left[-iwu\right]}\, g_0(u) = \frac{1}{2\pi} \exp{\left[-w^2/2\right]}.
\end{align}
Note that the time-dependent potential response of the system is nothing but $\Phi_{1k} = -(2\pi n_e e/\epsilon_0)(G_{1\kappa}(0)/k^2)$, where $G_{1\kappa}(0)$ is the inverse Laplace transform of $\Tilde{G}_{1\kappa}(0)$. We shall compute this in section~\ref{sec:combined_lin_resp}.

\section{Linear response: Continuum}\label{sec:cont_resp}

Computing the total response, $g_{1\kappa}(u,t)$, requires us to take the inverse Fourier transform in $u$ and the inverse Laplace transform in $t$ of equation~(\ref{G_1k_w}). As detailed in Appendix~A, first we substitute $\Tilde{G}_{1\kappa}(0)$ from equation~(\ref{G_1k_0}) in the expression for $\Tilde{G}_{1\kappa}(w)$ given in equation~(\ref{G_1k_w}), and then write it as a sum of three terms, $\Tilde{G}_{1\kappa}^{(1)}(w)$, $\Tilde{G}_{1\kappa}^{(2)}(w)$ and $\Tilde{G}_{1\kappa}^{(3)}(w)$. Performing the inverse Laplace transform of $\Tilde{G}_{1\kappa}(w)$ thus yields $g_{1\kappa}(u,t)$, which can be written as a sum of three terms that are the inverse Laplace transforms of $\Tilde{G}_{1\kappa}^{(1)}(w)$, $\Tilde{G}_{1\kappa}^{(2)}(w)$ and $\Tilde{G}_{1\kappa}^{(3)}(w)$ respectively: a continuum response, a discrete mode response (temporally convoluted with the continuum response), and a direct response to the perturber, also known as the `wake' (temporally convoluted with the discrete modes and the continuum). The continuum response depends on the continuous velocities of the particles, hence the name. The discrete mode response consists of coherent oscillations (modes) of the entire system. These oscillations occur at discrete frequencies, $\eta_n$ ($n=0,1,2,...$), which satisfy the dispersion relation, $D(\kappa,\eta_n)=0$. The wake response has a temporal dependence that is similar to that of the external perturber.

Let us first compute the continuum part of the response, $g_{1\kappa}^{\rm cont}(u,t)$, which amounts to computing the inverse Laplace transform of $\Tilde{G}_{1\kappa}^{(1)}(w)$ (first of equations~[\ref{G_1k_w_f_app}]). Assuming that the unperturbed DF is Maxwellian, this yields (see Appendix~\ref{App:lin_resp_calc} for details):

\begin{align}
g_{1\kappa}^{\rm cont}(u,t) &= \sqrt{\frac{2}{1-e^{-2\nu_\rmc t}}} \exp{\left[-\frac{u^2}{2\left(1-e^{-2\nu_\rmc t}\right)}\right]} \nonumber\\
&\times \exp{\left[-\kappa^2\left(\nu_\rmc t - 2\tanh{\frac{\nu_\rmc t}{2}}\right)\right]} \nonumber\\
&\times \exp{\left[-i \kappa u \tanh{\frac{\nu_\rmc t}{2}}\right]}\times \calF(\kappa,\nu_\rmc t),
\label{g1k_cont}
\end{align}
where

\begin{align}
&\calF(\kappa,\nu_\rmc t) \nonumber\\
&= \int_{-\infty}^\infty \rmd z \, \exp{\left[-{\left\{z+\left(\sqrt{2\tanh{\frac{\nu_\rmc t}{2}}}\kappa - i\frac{u}{\sqrt{2\left(1-e^{-2\nu_\rmc t}\right)}}\right)\right\}}^2\right]}\nonumber\\
&\times H_\kappa\left(\kappa+ze^{-\nu_\rmc t}\right).
\end{align}
Here, $H_\kappa$ is the velocity Fourier transform of the initial DF perturbation, $h_\kappa$.\\

\begin{figure*}
\centering
\includegraphics[width=1\textwidth]{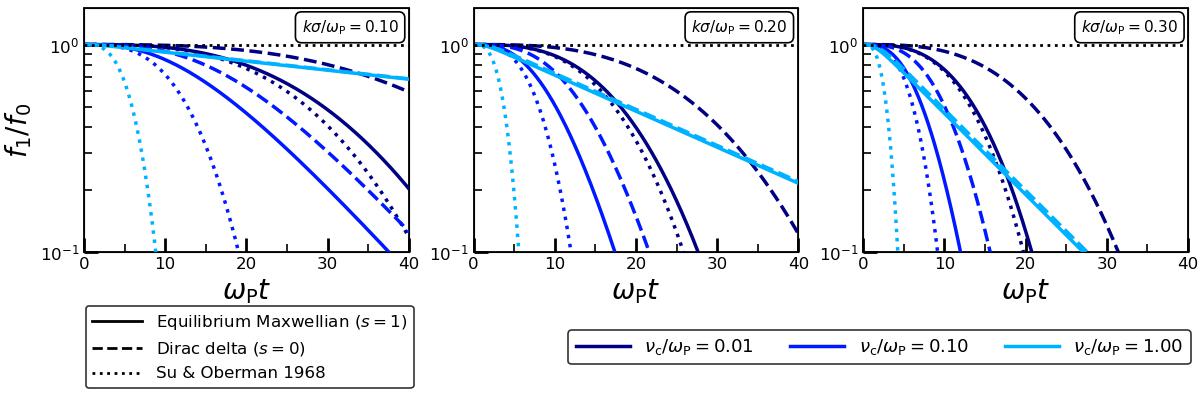}
\caption{The continuum response of a weakly collisional plasma as a function of time for $\nu_\rmc/\omega_\rmP = 0.01,0.1$ and $1$ indicated by the dark, medium dark and light blue lines, and three different values of $k\sigma$ in the three panels. The solid lines indicate the response when the initial perturbation in the DF is of the equilibrium Maxwellian form ($s=\sigma_0/\sigma = 1$). The dashed lines denote the response for an initial perturbation that is Dirac delta in the velocities ($s=0$). The dotted lines indicate the continuum response predicted by \cite{Su.Oberman.68}. Note that the $s=1$ and $s=0$ cases are very similar for $\nu_\rmc > k\sigma$, but different, albeit with the same temporal scaling, otherwise. The continuum response decays as $\sim \exp{\left[-{\left(k\sigma\right)}^2\nu_\rmc t^3/3\right]}$ for the $s=1$ case and $\sim \exp{\left[-{\left(k\sigma\right)}^2\nu_\rmc t^3/12\right]}$ for the $s=0$ case at $t<1/\nu_\rmc$, but switches over to a much slower $\sim \exp{\left[-\left({\left(k\sigma\right)}^2/\nu_\rmc\right)t\right]}$ decay at $t>1/\nu_\rmc$, unlike \cite{Su.Oberman.68} who only predict the former cubic exponential decay. Note that the collisional damping is most pronounced for $\nu_\rmc \sim k\sigma$, or in other words, when the perturbation scale, $\lambda \sim 1/k$, is comparable to the mean free path of collisions, $\lambda_\rmc = \sigma/\nu_\rmc$.}
\label{fig:f1_vs_t}
\end{figure*}

\subsection{Different models for $h_{\kappa}(u)$}\label{sec:init_pulse_models}

Evidently, the temporal behaviour of the continuum response depends on the functional form of the initial perturbation, $h_\kappa(u)$. We compute the response for a few specific models of $h_\kappa(u)$.

\subsubsection{Gaussian pulse}

Let us consider the case of an initial perturbation in the DF that is Gaussian or Maxwellian in velocities, but with a velocity dispersion different from the equilibrium velocity dispersion $\sigma$:

\begin{align}
h_\kappa(u) = \frac{h_{\kappa 0}}{\sqrt{2\pi s^2}}\exp{\left[-\frac{{\left(u-u_0\right)}^2}{2s^2}\right]}.
\label{initial_hk}
\end{align}
Here, $u_0$ is the drift velocity of the initial pulse, and $\sigma_0$ is its velocity dispersion, with $s = \sigma_0/\sigma$. The velocity Fourier transform of $h_\kappa(u)$ is given by

\begin{align}
H_\kappa(w) = \frac{h_{\kappa 0}}{2\pi} \exp{\left[-\frac{s^2w^2}{2}-i u_0 w \right]}.
\end{align}
In this case, the continuum response in equation~(\ref{g1k_cont}) becomes

\begin{align}
g_{1\kappa}^{\rm cont}(u,t) &= \frac{h_{\kappa 0}}{\sqrt{2\pi \delta^2_u(t)}} \exp{\left[-\frac{{\left(u-u_0e^{-\nu_\rmc t}\right)}^2}{2\delta^2_u(t)}\right]} \nonumber\\
&\times \exp{\left[-\kappa^2\delta^2_x(t)\right]}
\times \exp{\left[-i \kappa \left(\zeta^{(1)}_x(t)u+\zeta^{(2)}_x(t)u_0\right) \right]},
\label{g1k_cont_init_gaussian}
\end{align}
where

\begin{align}
&\zeta_x^{(1)}(t)
= \frac{\left(1-e^{-\nu_\rmc t}\right)\left(1-\left(1-s^2\right)e^{-\nu_\rmc t}\right)}{1-\left(1-s^2\right)e^{-2\nu_\rmc t}}, \nonumber\\
&\zeta_x^{(2)}(t) = \frac{{\left(1-e^{-\nu_\rmc t}\right)}^2}{1-\left(1-s^2\right)e^{-2\nu_\rmc t}},\nonumber\\
&\delta^2_u(t) = 1-\left(1-s^2\right)e^{-2\nu_\rmc t},\nonumber\\
&\delta^2_x(t) = \nu_\rmc t - \frac{1-e^{-\nu_\rmc t}}{2}\frac{3-e^{-\nu_\rmc t} + \left(1-s^2\right)\left(1-3 e^{-\nu_\rmc t}\right)}{1-\left(1-s^2\right)e^{-2\nu_\rmc t}}.
\end{align}

Let us analyze separately the key features of the above response. The first factor in equation~(\ref{g1k_cont_init_gaussian}) denotes the damping and broadening of the initial perturbation in the velocity space due to the collisional diffusion of particle velocities. The perturbation starts out being a Gaussian (in $u$) with a velocity dispersion $s$ (in units of $\sigma$), then gradually damps away and spreads out to attain the equilibrium Maxwellian form with velocity dispersion $\sigma$. 

\begin{figure}
\centering
\includegraphics[width=1\textwidth]{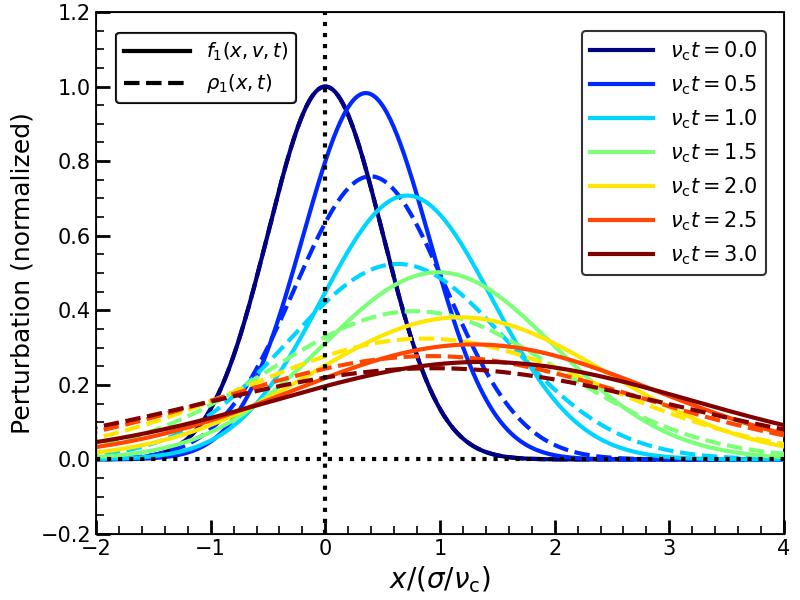}
\caption{The continuum response of a weakly collisional plasma as a function of $x$ (in units of $\sigma/\nu_\rmc$). The solid lines denote $f_1(x,v,t)$ (computed using equation~[\ref{f1_cont_gx_ddv}] and normalized by its maximum value at $t=0$) at seven different values of $\nu_\rmc t$ (ranging from $0$, indicated by the blue/uppermost lines, to $3$, indicated by the red/lowermost lines), while the dashed lines denote $\rho_1(x,t)$ (computed using equation~[\ref{rho1_cont_init_gaussian}] and normalized by its maximum value at $t=0$). We adopt $\sigma_{x,0}=0.5\sigma/\nu_\rmc$, $v=0.5\sigma$, $v_0=\sigma$ and $s=\sigma_0/\sigma=0.9$, i.e., the initial DF perturbation is close to the equilibrium Maxwellian form. Both quantities have been normalized by their maximum value at $t=0$. Note that both perturbations decay, broaden and slow down in $x$ due to collisional diffusion followed by damping, but the density perturbation decays faster due to the additional effect of phase-mixing.}
\label{fig:f1_rho1_vs_x}
\end{figure}

The second factor in equation~(\ref{g1k_cont_init_gaussian}) indicates the damping of the response (in $\kappa$) due to collisions. The extent of damping depends on the wave-number $k$ of the perturbation. The asymptotic temporal behaviour of this damping factor for the $s\neq 0$ case (we shall discuss the $s=0$ case, which corresponds to an initial Dirac delta pulse, separately) is as follows:

\begin{widetext}
\begin{align}
\exp{\left[-\kappa^2\delta^2_x(t)\right]} \approx
\begin{cases}
\exp{\left[-\kappa^2{\left(\nu_\rmc t\right)}^3/3\right]} = \exp{\left[-{\left(k\sigma\right)}^2\nu_\rmc t^3/3\right]},\;\;\;\;\;\;\;\,\nu_\rmc t\ll 1,\\
\exp{\left[-\kappa^2\nu_\rmc t\right]} = \exp{\left[-\left({\left(k\sigma\right)}^2/\nu_\rmc\right) t\right]},\;\;\;\;\;\;\;\;\;\;\;\;\;\;\nu_\rmc t\gg 1.
\end{cases}
\label{g1k_cont_asymptotic_damping_1}
\end{align}
\end{widetext}
The small time ($t\ll 1/\nu_\rmc$) behaviour of the above collisional damping factor is similar to the classic \citep[][]{Su.Oberman.68} result. This damping occurs cubic exponentially over a timescale, $\tau_{\rmd,1} = {\left(3/\nu_\rmc {\left(k\sigma\right)}^2\right)}^{1/3}$. However, the long time ($t\gg 1/\nu_\rmc$) damping occurs exponentially, much slower than the \cite{Su.Oberman.68} prediction. The timescale of this late time decay is $\tau_{\rmd,2} = \nu_\rmc/{\left(k\sigma\right)}^2$. When $\nu_\rmc \sim k\sigma$, or in other words the scale of the perturbation, $\lambda \sim 1/k$, is comparable to the mean free path, $\lambda_\rmc = \sigma/\nu_\rmc$, $\tau_{\rmd,1}$ is comparable to $\tau_{\rmd,2}$, and the corresponding response damps the most. This is evident in Fig.~\ref{fig:f1_vs_t}, which plots the amplitude of the continuum response as a function of time for three different values of $\nu_\rmc$ and three different values of $k\sigma$ as indicated. Our result, computed using equation~[\ref{g1k_cont_init_gaussian}], is shown by the solid and dashed lines for the $s=1$ and $0$ cases respectively, and the corresponding \cite{Su.Oberman.68} prediction is shown by dotted lines. The notable difference between our prediction and that of \cite{Su.Oberman.68} is a consequence of the fact that they neglected the damping/drag term in the LB operator, i.e., the first term on the RHS of equation~(\ref{lin_CBE_act_ang}). This approximation is justified only for $t\lesssim 1/\nu_\rmc$, when collisions cause the particle velocities to diffuse/disperse rather than damp on an average. This is the diffusive regime of collisions. However, collisions not only disperse the velocities of the particles but also damp out their individual mean velocities over time. This damping regime, which takes over at $t\gtrsim 1/\nu_\rmc$, is described by the first term on the RHS of equation~(\ref{lin_CBE_act_ang}), the very term that \citep[][]{Su.Oberman.68} neglect. By neglecting the damping term, \cite{Su.Oberman.68} have limited themselves to the diffusive regime of collisions, and therefore missed out on the crucial switchover of the decay to a slow exponential one in the long time damping regime. To obtain more physical insight into this, we shall discuss the collisional dynamics in the diffusive and damping regimes in detail, in section~\ref{sec:Langevin_analysis}, using a stochastic analysis with the Langevin equation, which is complementary to the Fokker-Planck approach taken so far.

The third factor in equation~(\ref{g1k_cont_init_gaussian}) is a phase factor that denotes the free streaming of the particles. It can be seen that at small time ($\nu_\rmc t \ll 1$), this factor evolves as $\exp{\left[-i\kappa u \nu_\rmc t\right]} = \exp{\left[-ikv t\right]}$. This behaviour is the same as that in the collisionless ($\nu_\rmc \to 0$) case. At large time ($\nu_\rmc t \gg 1$), the phase factor becomes $\exp{\left[-i\kappa \left(u+u_0\right)\right]}$, and therefore time-independent. This implies that the pulse comes to a halt after roughly the collision time, $1/\nu_\rmc$, owing to the constant drainage of linear momentum from a particle via collisions. This is a generic feature of collisional damping in Fokker-Planck dynamics.

So far we have seen that each $k$ mode of the continuum response damps away and decelerates due to collisional diffusion and damping. The spatial response is an integrated response of all $k$ modes. Since each $k$ mode damps away at a rate that depends non-linearly on $k$, the spatial response, which is the sum of all $k$ modes, will deviate from its initial functional form over time, i.e., the weakly collisional plasma acts as a dispersive medium. We shall demonstrate this by computing the spatial response for a given spatial distribution of the initial perturbation, i.e., a given $h_{\kappa 0}$. Since the response for each $k$ mode is Gaussian in $k$, as evident from equation~(\ref{g1k_cont_init_gaussian}) (for a Gaussian velocity distribution of the initial perturbation as we consider here), the integration over $k$ is straightforward for a Gaussian (in $k$) functional form of $h_{\kappa 0}$, i.e., a Gaussian (in $x$) functional form of the initial perturbation. The spatial response can be obtained by taking the inverse spatial Fourier transform of equation~(\ref{g1k_cont_init_gaussian}). For an initial Gaussian pulse with spatial dispersion $\sigma_{x,0}$, i.e., $g_1(x,u,0)=\left(F/{\sqrt{2\pi}}\right)\exp{\left[-x^2/2\sigma^2_{x,0}\right]}\,\exp{\left[-{\left(u-u_0\right)}^2/2s^2\right]}$, for which $h_{\kappa 0}=\left(F/2\pi\right)\exp{\left[-k^2\sigma^2_{x,0}/2\right]}\,\exp{\left[-{\left(u-u_0\right)}^2/2s^2\right]}$, we take the inverse Fourier transform of equation~(\ref{g1k_cont_init_gaussian}) and re-establish the dimensions to obtain the following form for the response:

\begin{widetext}
\begin{align}
&f_1^{\rm cont}(x,v,t) = \frac{n_e}{\sigma}\int_{-\infty}^\infty \rmd k\, \exp{\left[ikx\right]}\, g_{1\kappa}^{\rm cont}(u,t) \nonumber\\
&= \frac{n_e F}{\sqrt{2\pi\sigma^2\delta^2_u(t)}} \exp{\left[-\frac{{\left(v-v_0 e^{-\nu_\rmc t}\right)}^2}{2\sigma^2\delta^2_u(t)}\right]} \times \frac{1}{\sqrt{2\pi\left(\sigma^2_{x,0}+2{\left(\sigma/\nu_\rmc\right)}^2\delta^2_x(t)\right)}} \exp{\left[-\frac{{\left[x-\frac{1}{\nu_\rmc}\left(\zeta_x^{(1)}(t)v+\zeta_x^{(2)}(t)v_0\right)\right]}^2}{{2\left(\sigma^2_{x,0} + 2\left(\sigma/\nu_\rmc\right)^2\delta^2_x(t)\right)}}\right]}.
\label{f1_cont_gx_ddv}
\end{align}
\end{widetext}
Conservation of particle number ensures that the pulse damps out as well as broadens due to collisions, as manifested by both $x$ and $v$ parts of the above response. The spatial broadening of the pulse occurs as $\sim \left(2\sigma^2/3\nu_\rmc^2\right){\left(\nu_\rmc t\right)}^3$ for $t\lesssim 1/\nu_\rmc$, and as $\sim \left(2\sigma^2/\nu_\rmc^2\right)\nu_\rmc t$ for $t\gtrsim 1/\nu_\rmc$, while the spatial damping occurs as the inverse of this. If the width of the initial pulse is small, i.e., $\sigma_{x,0} \ll \sigma/\nu_\rmc$, it damps and broadens in the former way for a long time, i.e., we are in the diffusive regime. The typical damping timescale in this regime is $\sim {\left(3\sigma^2_{x,0}/2\sigma^2\nu_\rmc\right)}^{1/3}$. On the other hand, a well-extended initial pulse with $\sigma_{x,0} \gg \sigma/\nu_\rmc$, falls in the damping regime, i.e., it quickly transitions to the latter phase of exponential damping and broadening, which occurs over a relatively long timescale of $\sim \left(\sigma^2_{x,0}/2\sigma^2\right) \nu_\rmc$. Since both timescales increase with $\sigma_{x,0}$ (for fixed $\nu_\rmc$ and $\sigma$), we see that small-scale perturbations collisionally damp away faster than large-scale ones. Moreover, since the former timescale decreases with $\nu_\rmc$ (for fixed $\sigma_{x,0}$ and $\sigma$) and the latter increases, perturbations damp slowly in both small and large $\nu_\rmc$ limits. The damping and broadening of the pulse due to collisions is visible in Fig.~\ref{fig:f1_rho1_vs_x}, which plots $f_1$ (solid lines) as a function of $x$ at different points of time as indicated. Note that the pulse slows down over time due to collisional damping. The three key features of Fokker-Planck collisional dynamics are the damping, broadening and deceleration of a pulse, all three of which are manifest in Fig.~\ref{fig:f1_rho1_vs_x}.

Let us now study the evolution of the density perturbation, $\rho_1^{\rm cont}(x,t) = \int \rmd v\, f_1^{\rm cont}(x,v,t)$. Upon integrating equation~(\ref{f1_cont_gx_ddv}) over velocity, we obtain

\begin{align}
&\rho_1^{\rm cont}(x,t) \nonumber\\
&= \frac{n_e F}{\sqrt{2\pi\left[\sigma^2_{x,0}+{\left(\frac{\sigma}{\nu_\rmc}\right)}^2\left\{2\delta^2_x(t) + {\zeta^{(1)}_x}^2(t)\,\delta^2_u(t)\right\}\right]}} \nonumber\\
&\times \exp{\left[-\frac{{\left(x-\frac{v_0}{\nu_\rmc}\left(1-e^{-\nu_\rmc t}\right)\right)}^2}{2\left[\sigma^2_{x,0}+{\left(\frac{\sigma}{\nu_\rmc}\right)}^2\left\{2\delta^2_x(t) + {\zeta^{(1)}_x}^2(t)\,\delta^2_u(t)\right\}\right]}\right]}.
\label{rho1_cont_init_gaussian}
\end{align}
The density perturbation initially propagates at velocity $v_0$ and comes to a halt after $t\sim 1/\nu_\rmc$. At the same time, it damps and spreads out. This is not only due to collisional damping but also phase-mixing, as long as the initial perturbation has some finite spread in velocities, or in other words particles are initially perturbed over a finite velocity range. The width of the Gaussian scales as $\sim \sqrt{\sigma^2_{x,0} + \sigma^2 t^2}$ at small $t$, due to phase-mixing. At intermediate time, the width scales as $\sim \sqrt{\sigma^2_{x,0}+\sigma^2\nu_\rmc t^3/3}$, which is an outcome of the collisional diffusion of particle velocities. At late time, it scales as $\sim \sqrt{\sigma^2_{x,0} + 2\left(\sigma^2/\nu_\rmc\right) t}$, due to collisional damping. Due to the additional contribution of phase-mixing to the damping besides collisions, the density perturbation $\rho^{\rm cont}_1$ damps away faster than the fine-grained DF $f^{\rm cont}_1$, as manifest in Fig.~\ref{fig:f1_rho1_vs_x} that plots $f^{\rm cont}_1$ (solid lines) and $\rho^{\rm cont}_1$ (dashed lines) as functions of $x$.
\\
\subsubsection{Dirac delta pulse}\label{sec:init_pulse_Dirac_delta}

We now evaluate the response in the $s\to 0$ limit of the initial Gaussian pulse. This corresponds to an initial pulse that is Dirac delta in the particle velocities, i.e., $h_\kappa(u)= h_{\kappa 0}\delta(u-u_0)$, whose Fourier transform is $H_\kappa(w)=h_{\kappa 0}\exp{\left[-iu_0 w\right]}/(2\pi)$. In this case, the response is given by equation~(\ref{g1k_cont_init_gaussian}), with

\begin{align}
&\zeta_x^{(1)}(t)
= \zeta_x^{(2)}(t) = \tanh{\frac{\nu_\rmc t}{2}},\nonumber\\
&\delta^2_u(t) = 1-e^{-2\nu_\rmc t},\nonumber\\
&\delta^2_x(t) = \nu_\rmc t - 2\tanh{\frac{\nu_\rmc t}{2}}.
\label{g1k_cont_init_Dirac_delta}
\end{align}
The asymptotic temporal behaviour of the collisional damping factor is:

\begin{widetext}
\begin{align}
\exp{\left[-\kappa^2\delta^2_x(t)\right]} \approx
\begin{cases}
\exp{\left[-\kappa^2{\left(\nu_\rmc t\right)}^3/12\right]} = \exp{\left[-{\left(k\sigma\right)}^2\nu_\rmc t^3/12\right]},\;\;\;\nu_\rmc t\ll 1\\
\exp{\left[-\kappa^2\nu_\rmc t\right]} = \exp{\left[-\left({\left(k\sigma\right)}^2/\nu_\rmc\right) t\right]},\;\;\;\;\;\;\;\;\;\;\;\;\;\nu_\rmc t\gg 1.
\end{cases}
\label{g1k_cont_asymptotic_damping_2}
\end{align}
\end{widetext}
Note that the small time behaviour of the damping is the same as that in the general case given by equation~(\ref{g1k_cont_asymptotic_damping_1}), expect for the numerical factor of the exponent. The long time behaviour is exactly the same as equation~(\ref{g1k_cont_asymptotic_damping_1}).

The phase factor, $\exp{\left[-i\kappa(\zeta^{(1)})_x(t)u+\zeta^{(2)})_x(t)u_0\right]}$, scales as $\sim \exp{\left[-i\kappa(u+u_0)t/2\right]}=\exp{\left[-ik(v+v_0)t/2\right]}$ for $\nu_\rmc t\to 0$. In this limit, the velocity part of the response tends to $\delta(v-v_0)$, and thus the phase-factor goes to $\exp{\left[-ik v t\right]}$. This reproduces the form of the continuum response in the collisionless limit.

\subsubsection{Equilibrium Gaussian pulse}\label{sec:init_pulse_Maxwellian}

In the $s\to 1$ limit, the initial perturbation, $h_\kappa(u)$, is of the equilibrium Maxwellian form, i.e., $h_\kappa(u)=h_{\kappa 0}\exp{\left[-{\left(u-u_0\right)}^2/2\right]}/\sqrt{2\pi}$, and its Fourier transform is equal to $H_\kappa(w) = h_{\kappa 0}\exp{\left[-w^2/2-iu_0 w\right]}/(2\pi)$. In this case, the response is given by equation~(\ref{g1k_cont_init_gaussian}), with

\begin{align}
&\zeta_x^{(1)}(t) = 1-e^{-\nu_\rmc t},\nonumber\\
&\zeta_x^{(2)}(t) = {\left(1-e^{-\nu_\rmc t}\right)}^2,\nonumber\\
&\delta_u(t) = 1,\nonumber\\
&\delta^2_x(t) = \nu_\rmc t - \frac{\left(1-e^{-\nu_\rmc t}\right)\left(3-e^{-\nu_\rmc t}\right)}{2}.
\label{g1k_cont_init_eq_gaussian}
\end{align}
The phase factor is $\exp{\left[-ik\left(1-e^{-\nu_\rmc t}\right)\left(v+v_0\left(1-e^{-\nu_\rmc t}\right)\right)\right]}$, which scales as $\exp{\left[-ikvt\right]}$ in the $\nu_\rmc t \to 0$ limit. This recovers the form of the continuum response in the limit of zero collisions. The perturbation is initially of the equilibrium Maxwellian form and therefore remains so throughout, since the Maxwellian DF is an equilibrium solution of the Boltzmann-Poisson equations with the LB collision operator. The amplitude of the response, however, damps away due to collisional diffusion and damping. The damping factor has the same asymptotic behaviour as in the general case given by equation~(\ref{g1k_cont_asymptotic_damping_1}), i.e., the response damps away cubic exponentially till $t\sim 1/\nu_\rmc$ and transitions to a slower exponential decay afterwards.

\subsection{Langevin approach: diffusive and friction regimes}\label{sec:Langevin_analysis}

So far, we have seen that the collisional relaxation of a plasma occurs in two distinct regimes: the diffusive and friction regimes. The nature of relaxation in these two regimes is very different. These distinct behaviours of plasma relaxation can be understood by taking a closer look at the stochastic dynamics of collisions. We assume that the collisions are typically short-range random encounters, and therefore mimic a stochastic process that randomly changes the phase space coordinates of the particles. If the collisions are random enough, they erase the dynamical memory (initial conditions) of the colliding particles over time, i.e., the collisional scatterings can be thought of as Markovian random walks of the particles in phase space. A key consequence of this random (or chaotic) nature of collisions is the phenomenon of {\it molecular chaos}, which refers to the fact that the momenta of two colliding particles, despite being correlated immediately after a collision, become uncorrelated soon enough due to their encounters with other particles\footnote{The assumption of molecular chaos gives rise to the Boltzmann H theorem (second law of thermodynamics) in a system governed by short-range random/chaotic collisions, like the weakly collisional plasma described here.}. This justifies the use of the Boltzmann equation as the governing kinetic equation. This also ensures that the collisions are approximately uncorrelated in time and can be modelled by a stochastic acceleration with a white noise power spectrum. Hence, in the same spirit as that of modeling Brownian motion \citep[][]{Einstein.1905,Lemons.Gythiel.97}, we can model the collisional dynamics in a plasma using the Langevin equation, which is nothing but Newton's equation of motion with damping and stochastic forces:

\begin{align}
\frac{\rmd^2 x}{\rmd t^2} + \nu_\rmc \frac{\rmd x}{\rmd t} = \xi(t).
\label{Langevin_eq}
\end{align}
Here, $-\nu_\rmc \rmd x/\rmd t$ is the collisional damping/deceleration, also known as friction, and $\xi(t)$ is the stochastic acceleration due to collisions. We assume that $\xi(t)$ has a zero ensemble average and a white noise or uncorrelated power spectrum, i.e.,

\begin{align}
&\left<\xi(t)\right> = 0,\nonumber\\
&\left<\xi(t)\,\xi(t')\right> = \calD\,\delta(t-t'),
\label{white_noise}
\end{align}
where $\calD$ is a constant with the dimensions of velocity squared per unit time.

The Langevin equation is a stochastic differential equation which evolves the trajectory of each individual particle. The ensemble averages of the evolving phase space quantities and their dispersions, obtained by solving the Langevin equation, are the same as the first and second moments of these quantities, obtained with respect to the DF that evolves via the Fokker-Planck equation, as long as the initial DF perturbation is sufficiently close to the equilibrium Maxwellian form. In fact the Fokker-Planck collision operator corresponding to the Langevin equation given in equation~(\ref{Langevin_eq}) {\it is} the LB operator. The correspondence between the Langevin and Fokker-Planck equations therefore justifies that the results obtained in this section are an alternative way to obtain the various scalings of the response that are computed using the Fokker-Planck (LB) equation in the previous sections. 

Before solving the Langevin equation in its full glory, let us first take a careful look at the applicability of this equation in modeling plasma collisions. The key assumption is that the collisions can be modelled as short-range random encounters, despite the fact that Coulomb interactions between charged particles are long-range. The impact parameter of effective collisions is typically the Debye length, making the collisions functionally short-range. Hence, as long as collisional relaxation is dominated by scales larger than the Debye length ($\omega_\rmP \gtrsim k\sigma$), as is typically the case for a weakly collisional plasma, the collisions can be modelled as short-range stochastic encounters, and the Langevin (equivalently, the Fokker-Planck) description in this paper should properly capture the essential physics of plasma collisions. % (I think we should omit this decription here because there is a significant literature that we do not cite, and the problem has many subtleties that will take us off course). For $\omega_\rmP \lesssim k\sigma$, some of the collisions become long range. In this case, the stochastic acceleration in the Langevin equation~(\ref{Langevin_eq}) should explicitly depend on $x$ besides being time-dependent. The chaotic nature of collisions, which is the essence of the assumption of a white noise power spectrum for the collisions, holds as long as the number of particles within the Debye sphere, $n\lambda^3_\rmD$, is large, i.e., we are in the weakly-coupled plasma limit. If $n\lambda^3_\rmD$ is small, then collisions do not completely erase dynamical memory and therefore can be non-Markovian in nature. This would entail that the power spectrum of the stochastic collision term in equation~(\ref{Langevin_eq}) is red noise/correlated in time instead of being white noise/uncorrelated as we assume. To properly capture the long range and non-Markovian nature of collisions in the above cases, we would have to invoke in the kinetic equation a more sophisticated collision operator than the LB operator, such as the Landau or the Balescu-Lenard operator. We leave a detailed perturbative analysis using such collision operators for future work.%

With the aforementioned assumptions, let us now solve the Langevin equation to evaluate the mean trajectories of particles and the position and velocity dispersions around them. Integrating equation~(\ref{Langevin_eq}) once with respect to $t$, with the initial condition, $v(0)=v_0$, we obtain the following form for the velocity:

\begin{align}
v(t) = \frac{\rmd x}{\rmd t} = v_0\,e^{-\nu_\rmc t} + \int_0^t \rmd t'\,e^{-\nu_\rmc(t-t')} \xi(t').
\label{v_Langevin}
\end{align}
Since the ensemble average of $\xi(t)$ is zero, the ensemble averaged or mean velocity is equal to 

\begin{align}
\left<v(t)\right> = v_0\,e^{-\nu_\rmc t}.
\label{v_mean}
\end{align}
The exponential decay of the mean velocity manifests damping or friction: the constant drainage of momentum from each particle due to collisions. 

The velocity dispersion is given by (see Appendix~\ref{App:vel_pos_disp_Langevin_calc} for detailed derivation)

\begin{align}
\sigma^2(t) &= \left<\left(v(t)-v_0\,e^{-\nu_\rmc t}\right)^2\right> \nonumber\\
&= \frac{\calD}{\nu_\rmc} \left(1-e^{-2\nu_\rmc t}\right),
\label{vel_disp_Langevin}
\end{align}
where we have used the uncorrelated nature of the stochastic force, i.e., the second of equations~(\ref{white_noise}). Note that the velocity dispersion initially increases linearly with time, i.e., $\sigma^2(t)\sim 2\calD\, t$. This is the essence of velocity diffusion or particle random walk in the velocity space, and is caused by the stochastic acceleration, $\xi(t)$. The late-time dynamics, which sets in after $t\sim 1/\nu_\rmc$, is governed by the frictional deceleration, $-\nu_\rmc \rmd x/\rmd t$. This friction/damping drives $\sigma^2$ towards the equilibrium value, $\lim_{t\to\infty}\sigma^2(t)=\sigma^2$. Thus, from the above equation, we infer that $\calD = \nu_\rmc\sigma^2$, which is the celebrated Einstein relation. $\calD$ is the velocity diffusion coefficient, which is related to the damping coefficient or collision frequency, $\nu_\rmc$, by the proportionality factor, $\sigma^2$. Therefore, friction and diffusion go hand-in-hand during collisions, and one cannot exist without the other. Consequently, neglecting the friction term in the collision operator \citep[]{Su.Oberman.68,Short.Simon.02} provides an incomplete description of collisional dynamics, which may lead to erroneous conclusions about the long time decay of the response in weakly collisional plasmas.

Let us now evaluate the time dependence of the particle position. Integrating equation~(\ref{v_Langevin}) with respect to $t$, with the initial condition that $x(0)=0$, we obtain

\begin{align}
x(t) &= \frac{v_0}{\nu_\rmc}\left(1-e^{-\nu_\rmc t}\right) + \int_0^t \rmd t'\int_0^{t'} \rmd t''\,e^{-\nu_\rmc\left(t'-t''\right)} \xi(t'').
\label{x_Langevin}
\end{align}
Recalling that $\left<\xi(t)\right>=0$, we see that the mean position is equal to 

\begin{align}
\left<x(t)\right> = \frac{v_0}{\nu_\rmc}\left(1-e^{-\nu_\rmc t}\right),
\label{x_mean}
\end{align}
i.e., the particle gradually slows down and traverses an overall distance of $v_0/\nu_\rmc$. The mean-squared position is given by (see Appendix~\ref{App:vel_pos_disp_Langevin_calc} for detailed derivation)

\begin{align}
\sigma^2_x(t) &= \left<{\left[x(t)-\frac{v_0}{\nu_\rmc}\left(1-e^{-\nu_\rmc t}\right)\right]}^2\right> \nonumber\\
&= {\left(\frac{\sigma}{\nu_\rmc}\right)}^2 \left[\nu_\rmc t - \frac{\left(1-e^{-\nu_\rmc t}\right)\left(3-e^{-\nu_\rmc t}\right)}{2}\right],
\label{mean_sq_x_Langevin}
\end{align}
where we have used the second of equations~(\ref{white_noise}), i.e., $\xi(t)$ is a white noise source. Initially, in the diffusive regime, where the dynamics is predominantly driven by the stochastic acceleration $\xi(t)$, the mean-squared position scales as $\sigma^2_x(t) \sim \nu_\rmc \sigma^2 t^3/3$. This is a consequence of the fact that the velocity dispersion, $\sigma^2 \approx \rmd^2\sigma^2_x/\rmd t^2$, scales as $\sim 2 \calD\,t = 2\sigma^2 \nu_\rmc t$ in this regime. Finally, after $t\sim 1/\nu_\rmc$, the mean-squared position scales as $\sigma^2_x(t) \sim (\sigma^2/\nu_\rmc)\,t$. This owes to the fact that the velocity dispersion, $\sigma^2(t) \approx \nu_\rmc\, \rmd\sigma^2_x/\rmd t$, attains a nearly constant value, $\sigma^2$, due to friction. In this late-time friction-dominated regime, the velocity evolution is governed by the frictional deceleration, but the evolution of the mean-squared position is still driven by the stochastic acceleration, $\xi(t)$.

The temporal dependence of the mean and mean squared velocity and position obtained from the above analysis is very similar to what shows up in our analysis of plasma relaxation using the LB operator in the previous sections. The exponentially decaying drift velocity obtained from the Langevin analysis (equation~[\ref{v_mean}]), is the same as the drift velocity of the pulse obtained using the LB analysis (see equations~[\ref{g1k_cont_init_gaussian}] and \ref{f1_cont_gx_ddv}). The slowing drift of the mean position is manifest in both the Langevin (equation~[\ref{x_mean}]) and LB (equation~[\ref{rho1_cont_init_gaussian}]) analyses. The time-dependence of the velocity dispersion is also the same; compare $\sigma^2(t)$ in equation~(\ref{vel_disp_Langevin}) with $\sigma^2 \delta^2_u(t)$ in equation~(\ref{f1_cont_gx_ddv}), where $\delta^2_u(t)$ is given by the second of equations~(\ref{g1k_cont_init_Dirac_delta}). And the mean squared position, computed using the Langevin equation (equation~[\ref{mean_sq_x_Langevin}]) is the same as the spatial width of the response computed using the LB analysis (note $\delta^2_x(t)$ in equation~[\ref{g1k_cont_init_eq_gaussian}]). The prefactors of the above quantities slightly differ between the LB and Langevin analyses when the initial perturbation in the DF is very different from the equilibrium Maxwellian form. This is because the diffusion coefficients of the LB operator rely on the Einstein relation, and are therefore dictated by the equilibrium Maxwellian DF. The correspondence is exact only when the initial pulse is of the equilibrium Maxwellian form in velocities. Still, the asymptotic temporal behaviour of $\left<x\right>$, $\left<v\right>$, $\sigma^2_x$ and $\sigma^2$ obtained using the Langevin analysis is the same as that obtained using the LB analysis in all cases. This proves that the LB operator, despite being only an approximate description of collisional dynamics, properly captures the essential physics of collisions, particularly in the small and large time limits. Put differently, the diffusive and friction regimes of collisional relaxation predicted by the LB operator in section~\ref{sec:init_pulse_models} are not merely artefacts of the form of the collision operator chosen, but are qualitatively robust phenomena in a system governed by weak collisions.

\section{Linear response: discrete modes}\label{sec:normal_modes}

So far, we have studied the first part of the plasma response, the continuum, which encapsulates the collisionally damped streaming of particles. The second part of the response on the other hand consists of discrete modes\footnote{For each $k$, there are countably infinite modes.} that are collective coherent oscillations of the system. It can be computed by taking the inverse Laplace transform of $\Tilde{G}_{1\kappa}^{(2)}(\omega)$ in equation~(\ref{G_1k_w_f_app}). The resulting response involves a convolution of the continuum and discrete mode responses. We defer the computation of the combined response to the next section, and only discuss the behaviour of the discrete modes in this section. The discrete modes oscillate at frequencies that follow the dispersion relation,

\begin{align}
D(\kappa,\eta) &= 1+ {\left(\frac{\omega_\rmP}{\kappa\nu_\rmc}\right)}^2 \left[\,1-\left(a-\zeta\right)\zeta^{-a}\exp{\left[\zeta\right]}\,\gamma\left(a,\zeta\right)\,\right] = 0,
\label{plasma_disp_rel}
\end{align}
where $a=\eta+\kappa^2$, $\zeta = \kappa^2$, and $\gamma(a,\zeta)=\int_0^\zeta \rmd z\,z^{a-1}\exp{\left[-z\right]}$ is the lower incomplete gamma function. These frequencies are in general complex. The numerical computation of these frequencies is quite involved and the standard root-finding algorithms are found to be computationally expensive and imprecise. Therefore, we develop a novel root-finding algorithm to accomplish this. We rewrite the above dispersion relation as the following equations:

\begin{align}
&\exp{\left[\zeta\right]}\;{\rm Re}\left[\left(a-\zeta\right)\zeta^{-a}\,\gamma\left(a,\zeta\right)\right] = 1+{\left(\frac{\kappa\nu_\rmc}{\omega_\rmP}\right)}^2,\nonumber\\
&\exp{\left[\zeta\right]}\;{\rm Im}\left[\left(a-\zeta\right)\zeta^{-a}\,\gamma\left(a,\zeta\right)\right] = 0,
\end{align}
where ${\rm Re}$ and ${\rm Im}$ denote real and imaginary parts respectively. These equations are non-linear in ${\rm Re}(a)$ and ${\rm Im}(a)$, i.e., in ${\rm Re}(\eta)$ and ${\rm Im}(\eta)$. We simultaneously solve these non-linear equations in the following way. For each $\kappa$ and $\omega_\rmP/\nu_\rmc$, we compute two sets of contours in the ${\rm Re}(\eta)-{\rm Im}(\eta)$ plane. The first set of contours corresponds to those values of ${\rm Re}(\eta)$ and ${\rm Im}(\eta)$ that satisfy the first of the above equations. The second set corresponds to those values of ${\rm Re}(\eta)$ and ${\rm Im}(\eta)$ that satisfy the second equation. Now, we compute the intersection points of the first and second sets of contours, which are nothing but the simultaneous solutions of the above equations in the ${\rm Re}(\eta)$-${\rm Im}(\eta)$ space. The $\eta$ thus obtained, follow the dispersion relation given by equation~(\ref{plasma_disp_rel}), and are therefore the discrete mode frequencies.

\begin{figure}
\centering
\includegraphics[width=1\textwidth]{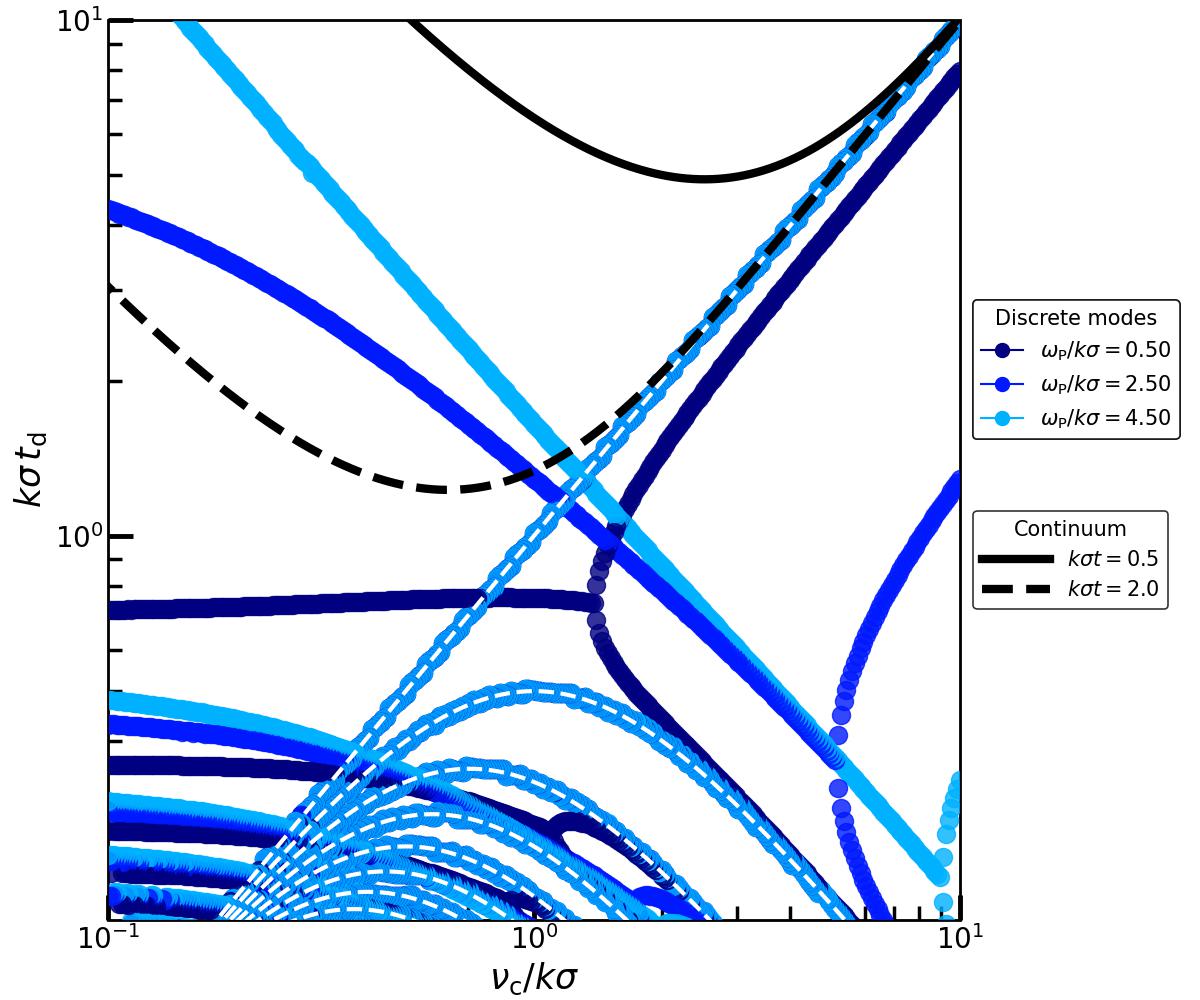}
\caption{The damping time, $t_\rmd$ (in units of $1/k\sigma$), of the discrete modes and the continuum as a function of $\nu_\rmc/k\sigma$. The dark, medium dark and light blue dots indicate the discrete modes for $\omega_\rmP/k\sigma=0.5,2.5$ and $4.5$ respectively. The white dashed lines denote the discrete modes of the Lenard-Bernstein (LB) branch. The rest belong to the Landau branch. The solid and dashed lines denote the damping time of the continuum response at $t=0.5$ and $2/k\sigma$ respectively. Note that at small time ($t\lesssim 1/\nu_\rmc$), the continuum response damps slower than the discrete modes. At later times, the discrete modes damp slower than the continuum for $\nu_\rmc<k\sigma$. For $\nu_\rmc>k\sigma$, the continuum response damps at the same rate as the zero-th order LB mode. The discrete modes of the Landau branch smoothly tend to the Landau modes in the $\nu_\rmc/k\sigma\to 0$ limit. At $\nu_\rmc \approx 2\omega_\rmP$, the Landau branch modes bifurcate and transition from being oscillatory (for $\nu_\rmc \lesssim 2\omega_\rmP$) to non-oscillatory (for $\nu_\rmc \gtrsim 2\omega_\rmP$).}
\label{fig:damping_time}
\end{figure}

%\begin{figure}
%\centering
%\includegraphics[width=0.95\textwidth]{osc_freq_vs_nuc_by_ksigma_complex_modes.jpeg}
%\caption{}
%\label{fig:osc_freq}
%\end{figure}

The discrete mode spectrum of a weakly collisional plasma has a rich structure. As shown by \citep[][etc.]{Landau.46,Lenard.Bernstein.58}, all modes in a plasma that is perturbed around an $f_0$ with $\partial f_0/\partial v < 0$ (e.g., the Maxwellian DF) are damped, i.e., have ${\rm Re}(\gamma)<0$. The damping time is given by $t_\rmd = 1/\left|{\rm Re}(\gamma)\right|$. Fig.~\ref{fig:damping_time} plots this damping time (in units of $1/k\sigma$), as a function of $\nu_\rmc/k\sigma$. The different colors denote different values of $\omega_\rmP/k\sigma$ as indicated. The white dashed lines trace out a particular family of modes called the Lenard-Bernstein or LB modes, which are non-oscillating (discussed in detail in section~\ref{sec:LB_modes}). The black solid and dashed lines indicate the damping time of the continuum response, defined as ${\left|\partial \ln{g_{1\kappa}^{\rm cont}}/\partial t\right|}^{-1}$, at $t=0.5$ and $2$ times $1/k\sigma$, respectively. Since the continuum response scales as $\sim \exp{\left[-{(k\sigma)}^2\nu_\rmc t^3/3\right]}$ at small time, and the modal response scales as $\sim \exp{\left[-\gamma t\right]}$ throughout, the continuum response dominates over the discrete modes at short times. At long times, the discrete modes dominate on scales smaller than the mean free path ($k>\nu_\rmc/\sigma$). However, on scales larger than the mean free path ($k<\nu_\rmc/\sigma$), the dominant modes are the continuum as well as the least damped discrete mode, which is one of the non-oscillating LB modes (see section~\ref{sec:LB_modes} for details).

The discrete modes in a weakly collisional plasma are diverse in nature and deserve a systematic categorization. We identify two distinct classes of discrete modes: (i) the Landau branch, which analytically continues to the Landau modes in the collisionless limit ($\nu_\rmc \ll k\sigma$) and are Landau damped, and (ii) the Lenard-Bernstein (LB) branch, which consists of an extra set of modes that appear in addition to the Landau-like modes for non-zero $\nu_\rmc$. We now discuss these two families of discrete modes in detail.
\\
\subsection{Landau branch}\label{sec:Landau_modes}

The modes belonging to the Landau branch analytically continue to the Landau modes in the collisionless ($\nu_\rmc/k\sigma\to 0$) limit. As evident from Fig.~\ref{fig:damping_time}, the damping timescales of the Landau branch modes tend towards the Landau values in the $\nu_\rmc/k\sigma\to 0$ limit, unlike the LB branch modes, indicated by the white dashed lines, whose damping times tend to zero. The analytic continuation to the Landau values can also be seen by rewriting the dispersion relation given in equation~(\ref{plasma_disp_rel}) in the following form \cite{Lenard.Bernstein.58}:

\begin{align}
D(k,\gamma) &= 1 + \frac{\omega^2_\rmP}{\nu_\rmc} \int_0^\infty \rmd s\, \left(1-\exp{\left[-\nu_\rmc s\right]}\right) \nonumber\\
&\times \exp{\left[-\left(\gamma+\frac{k^2\sigma^2}{\nu_\rmc}\right)s +\frac{k^2\sigma^2}{\nu_\rmc^2}\left(1-e^{-\nu_\rmc s}\right)\right]} = 0.
\label{plasma_disp_rel_mod}
\end{align}
In the $\nu_\rmc/k\sigma \to 0$ limit, this reduces to the Landau dispersion relation \citep[][]{Landau.46},

\begin{align}
D(k,\gamma) = 1 + \omega^2_\rmP \int_0^\infty \rmd s\,s \exp{\left[-\left(\gamma s + \frac{k^2\sigma^2 s^2}{2}\right)\right]} = 0.
\label{Landau_dispersion_relation}
\end{align}
The solutions of this equation are the Landau modes, $\gamma_0$, which are complex. For each $k$, there are infinitely many discrete modes\footnote{If the equilibrium DF is different from Maxwellian, e.g., if it is a Lorentzian in $v$, there can be finitely many discrete modes for each $k$.}. Of these, the least damped modes have the following asymptotic form \citep[][]{Landau.46}:

\begin{widetext}
\begin{align}
\gamma_0 &= -2k\sigma\sqrt{\ln{\left({k\sigma}/{\omega_\rmP}\right)}} \pm \frac{i\pi}{2} \frac{k\sigma}{\sqrt{\ln{\left({k\sigma}/{\omega_\rmP}\right)}}},\qquad \qquad \qquad \qquad \; \, k\sigma \gg \omega_\rmP,\nonumber\\
&= -\sqrt{\frac{\pi}{8}}\,k\sigma\, {\left(\frac{\omega_\rmP}{k\sigma}\right)}^4 \exp{\left[-\frac{1}{2}{\left(\frac{\omega_\rmP}{k\sigma}\right)}^2\right]} \pm i\omega_\rmP\left[1+\frac{3}{2}{\left(\frac{k\sigma}{\omega_\rmP}\right)}^2\right], \qquad \, k\sigma \ll \omega_\rmP.
\end{align}
\end{widetext}
The above form shows that, in the collisionless limit, the modes are oscillatory and very weakly damped on scales much larger than the Debye scale ($k\ll k_\rmD = \omega_\rmP/\sigma$). These are also known as Langmuir modes. On scales smaller than the Debye scale ($k\gg k_\rmD$), the modes are oscillatory but strongly (Landau) damped. Hence, oscillations in a collisionless plasma are damped by collective excitations only within the Debye scale.

The Landau modes are modified in presence of collisions, i.e., for non-zero $\nu_\rmc$. Let us compute the leading order correction to $\gamma$ in presence of collisions, using linear perturbation theory in the neighborhood of the Landau modes, $\gamma_0$ (taking $\nu_\rmc \ll \min{\left[\gamma_0,k\sigma\right]}$). We expand $\gamma$ about $\gamma_0$ as $\gamma=\gamma_0 + \nu_\rmc\, \delta_1$. Then, we rewrite the dispersion relation given in equation~(\ref{plasma_disp_rel_mod}) by only retaining terms that are zero-th and linear order in $\nu_\rmc$, as follows:

\begin{align}
&D(k,\gamma_0+\nu_\rmc\delta_1) \nonumber\\
&= 1 + \omega^2_\rmP \int_0^\infty \rmd s\, \left[s - \nu_\rmc s^2 \left(\delta_1 + \frac{1}{2} - \frac{k^2\sigma^2 s^2}{6}\right)\right] \nonumber\\
&\times \exp{\left[-\left(\gamma_0 s + \frac{k^2\sigma^2 s^2}{2}\right)\right]} = 0.
\end{align}
Using that $\gamma_0$ follow the Landau dispersion relation (equation~[\ref{Landau_dispersion_relation}]), and performing the $s$ integrals, we obtain:

\begin{align}
\delta_1 = \frac{\gamma-\gamma_0}{\nu_\rmc} = -\frac{1}{2} + \frac{1}{6} \frac{P(\gamma_0/k\sigma)}{Q(\gamma_0/k\sigma)},
\end{align}
where

\begin{align}
&P\left(\frac{\gamma_0}{k\sigma}\right) = \left[{\left(\frac{\gamma_0}{k\sigma}\right)}^3 + 4\frac{\gamma_0}{k\sigma}\right]\nonumber\\
&+\, \left[{\left(\frac{\gamma_0}{k\sigma}\right)}^4+6{\left(\frac{\gamma_0}{k\sigma}\right)}^2+3\right]\sqrt{\frac{\pi}{2}}\left[1+{\rm erf}\left(\frac{\gamma_0}{k\sigma}\right)\right]\exp{\left[\frac{1}{2}{\left(\frac{\gamma_0}{k\sigma}\right)}^2\right]},\nonumber\\
&Q\left(\frac{\gamma_0}{k\sigma}\right) = \frac{\gamma_0}{k\sigma}\nonumber\\
&+ \,\left[{\left(\frac{\gamma_0}{k\sigma}\right)}^2+1\right]\sqrt{\frac{\pi}{2}}\left[1+{\rm erf}\left(\frac{\gamma_0}{k\sigma}\right)\right]\exp{\left[\frac{1}{2}{\left(\frac{\gamma_0}{k\sigma}\right)}^2\right]}.
\end{align}
On scales much larger than the Debye length ($k\sigma \ll \omega_\rmP$), $\gamma_0/k\sigma\to 0$, and therefore, $P/Q\to 0$. This implies that $\delta_1 \to -\frac{1}{2}$, and therefore

\begin{align}
\gamma \approx \gamma_0-\frac{\nu_\rmc}{2}
\end{align}
on large scales. In Fig.~\ref{fig:damping_time}, this is evident from the linear decrease with $\nu_\rmc$ (at large $\nu_\rmc$) of the damping time of the most weakly damped modes for the $\omega_\rmP/k\sigma=4.5$ and $2.5$ cases, indicated by the light and medium dark blue dots respectively.

Let us now investigate the case of frequent collisions, i.e., when $\nu_\rmc \gtrsim k\sigma$. As evident from Fig.~\ref{fig:damping_time}, the modes remain oscillatory but become more strongly damped with increasing $\nu_\rmc$, until a bifurcation occurs and they become non-oscillatory. At bifurcation, the complex frequencies of the modes become purely real and negative. The bifurcation occurs when $\nu_\rmc \sim 2\omega_\rmP$, i.e., when the mean free path, $\lambda_\rmc = \sigma/\nu_\rmc$, is roughly half the Debye length, $\lambda_\rmD = \sigma/\omega_\rmP$. Each bifurcation results in a less damped upper branch and a more damped lower branch. For the zeroth order modes (which analytically continue to the least damped Landau modes in the $\nu_\rmc/k\sigma \to 0$ limit), the damping timescale of the upper branch scales as $t_\rmd \sim \nu_\rmc$ while that of the lower branch scales as $t_\rmd \sim 1/\nu_\rmc$ in the $\nu_\rmc \gg 2\omega_\rmP$ limit. Hence, the zero-th order Landau branch mode is very weakly damped, albeit non-oscillatory for $\nu_\rmc \gg 2\omega_\rmP$. For the higher order modes, the damping timescales of both branches scale as $t_\rmd \sim 1/\nu_\rmc$, i.e., these modes are strongly damped as well as non-oscillatory in the $\nu_\rmc \gg 2\omega_\rmP$ limit.

The above behaviour of the modes at large $\nu_\rmc/2\omega_\rmP$ can be readily seen by analyzing the dispersion relation in the $\nu_\rmc/k\sigma\to \infty$ limit. We follow \citep[][]{Lenard.Bernstein.58} to obtain the following form for the dispersion relation (for the zero-th order mode) in this limit:

\begin{align}
&D(k,\gamma) = 1 + {\left(\frac{\omega_\rmP}{\nu_\rmc}\right)}^2 \int_0^\infty \rmd x\,x^{\gamma/\nu_\rmc - 1}(1-x) = 0,
\end{align}
which can be integrated to yield

\begin{align}
1 + {\left(\frac{\omega_\rmP}{\nu_\rmc}\right)}^2 \left[\frac{1}{\gamma/\nu_\rmc} - \frac{1}{\gamma/\nu_\rmc + 1}\right] = 0.
\end{align}
This is a quadratic equation in $\gamma/\nu_\rmc$ and has the following solutions:

\begin{align}
\frac{\gamma}{\nu_\rmc} = -\frac{1}{2}\left[1\pm \sqrt{1-{\left(\frac{2\omega_\rmP}{\nu_\rmc}\right)}^2}\right].
\end{align}
The frequencies are complex conjugate, i.e., the zero-th order modes are damped but oscillating for $\omega_\rmP>\nu_\rmc/2$. At $\omega_\rmP=\nu_\rmc/2$, a bifurcation occurs and the frequencies become negative real, i.e., the modes become non-oscillatory and damped. For $\nu_\rmc \gg 2\omega_\rmP$, the upper branch mode has $\gamma \approx -\omega^2_\rmP/\nu_\rmc$ and is weakly damped, while the lower branch one has $\gamma \approx -\nu_\rmc$ and is strongly damped.

From Fig.~\ref{fig:damping_time} as well as the above analysis, we see that the discrete modes are more strongly damped than the Landau modes of the collisionless limit when $\nu_\rmc < 2\omega_\rmP$. At $\nu_\rmc \approx 2\omega_\rmP$, each of these modes bifurcates into two non-oscillatory modes, one of which is less damped than the other. The less damped mode eventually becomes more weakly damped than the corresponding Landau mode at $\nu_\rmc \gg 2\omega_\rmP$. Therefore, collisions facilitate the collective, Landau-like damping of plasma perturbations for $\nu_\rmc < 2\omega_\rmP$, i.e., when the mean free path, $\lambda_\rmc = \sigma/\nu_\rmc$, exceeds half the Debye length, $\lambda_\rmD = \sigma/\omega_\rmP$. On the other hand, for $\nu_\rmc \gg 2\omega_\rmP$, i.e., $\lambda_\rmc \ll \lambda_\rmD/2$, collisions oppose Landau damping, but make the modes non-oscillatory. Since the mean free path is smaller than the Debye length in this case, collisional scatterings destroy the coherence of the collective oscillations.

\begin{figure*}
\centering
\includegraphics[width=0.95\textwidth]{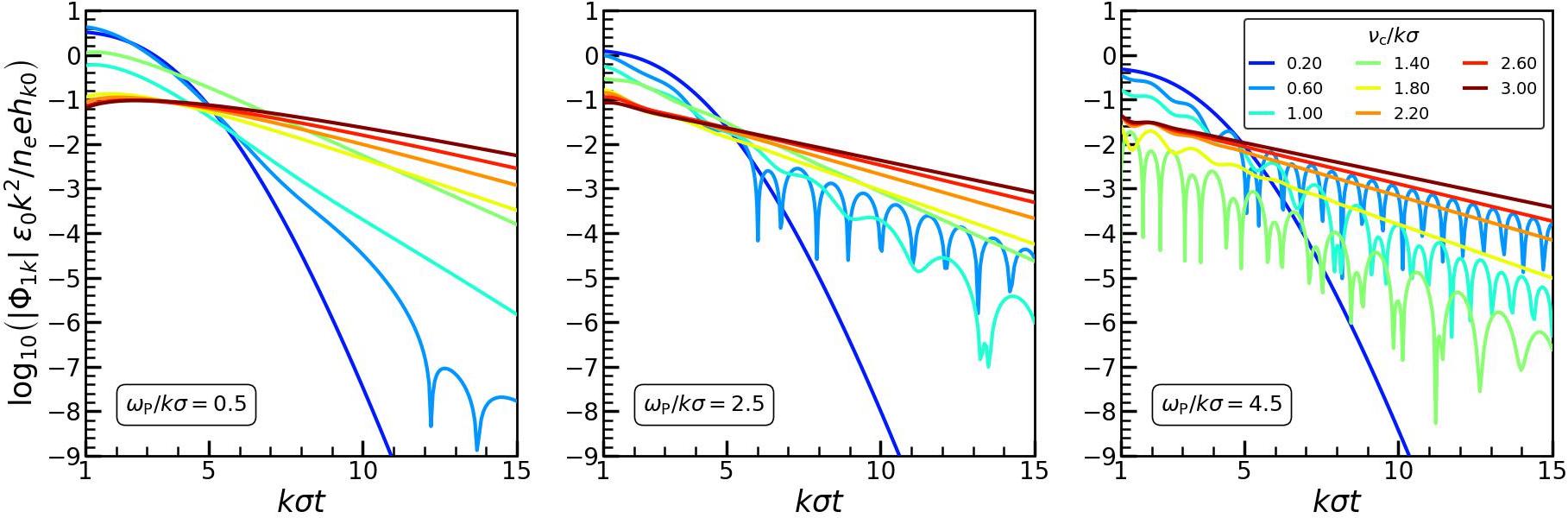}
\caption{The combined temporal response of a weakly collisional plasma, including both the continuum and the discrete modes, in terms of $\left|\Phi_{1k}\right|$ vs $t$ (in units of $1/k\sigma$). The initial perturbation in the DF is taken to be a Dirac delta function in $v$ ($s=\sigma_0/\sigma=0$). Different colors indicate the response for different values of $\nu_\rmc/k\sigma$, and different panels correspond to different values of $\omega_\rmP/k\sigma$ as indicated. The initial phase of the decay, up to $t\sim 1/\nu_\rmc$, is driven by the continuum. For $\nu_\rmc<k\sigma$, the latter phase of the decay is driven by the discrete modes, while for $\nu_\rmc>k\sigma$, the long term decay is governed by the continuum and the zero-th order LB mode, which is a non-oscillatory mode.}
\label{fig:Phi1k_vs_t}
\end{figure*}

\subsection{Lenard-Bernstein Branch}\label{sec:LB_modes}

The collision operator in the Boltzmann equation involves the highest velocity derivative and is therefore a singular perturbation to the collisionless Boltzmann equation. This observation foreshadows our finding that the discrete mode spectrum for the weakly collisional case is qualitatively different from the collisionless case. Therefore, in addition to the Landau branch modes discussed above, the LB operator introduces an additional set of modes, which we term Lenard-Bernstein or LB modes. We indicate these using white dashed lines in Fig.~\ref{fig:damping_time}. Unlike the Landau branch modes, the LB modes do not analytically continue to the Landau modes in the collisionless limit. Moreover, the LB modes are damped and non-oscillatory. As also found by \citep[][]{Lenard.Bernstein.58}, the frequencies of these modes are given by

\begin{align}
\frac{\gamma}{\nu_\rmc} = -{\left(\frac{k\sigma}{\nu_\rmc}\right)}^2 - n,
\end{align}
where $n = 0,1,2,...\,$. These modes are independent of the plasma frequency and therefore completely decoupled from the mean electrostatic field. These solely depend on $\nu_\rmc$, and are thus pure manifestations of short range collisional scatterings in the plasma. The $n\neq 0$ modes are heavily damped, especially in large and small $\nu_\rmc/k\sigma$ limits, since $\gamma\approx -n\nu_\rmc$ for $\nu_\rmc \gg k\sigma$ and $\gamma \approx -{\left(k\sigma\right)}^2/\nu_\rmc$ for $\nu_\rmc \ll k\sigma$. On the other hand, the $n=0$ mode is heavily damped for small $\nu_\rmc/k\sigma$ but very weakly damped in the large $\nu_\rmc/k\sigma$ limit, since $\gamma =-{\left(k\sigma\right)}^2/\nu_\rmc$ throughout. 

The LB modes play a crucial role in plasma relaxation in the large $\nu_\rmc/k\sigma$ limit. The damping timescale of the LB modes scales with $\nu_\rmc$ in the same way as that of the Landau branch modes in this limit. Also, in this limit, the damping time of the continuum response is the same as that of the $n=0$ LB mode. Hence, it is the LB modes that dictate the relaxation of plasmas in the limit of frequent collisions, or equivalently, on scales larger than the mean free path.

Vlasov simulations have demonstrated some of the phenomenology of wave damping described in this paper. \cite{Banks.etal.16}  find in their simulation the survival of a non-oscillating and slowly decaying `entropy mode' (also discussed by \citep{Brantov.Bychenkov.12}) in the long-time limit. This bears qualitative resemblance to the LB modes \cite{Lenard.Bernstein.58} discussed in this paper. The long term, slow decay of the response in their simulation is very similar to that of the slowly decaying continuum and the $n=0$ LB mode that we predict (although non-linear Landau damping effects become important after approximately the bounce time or libration time for the wave-particle resonances). The pronounced damping of the response for $\nu_\rmc \sim k\sigma$, as predicted by our analysis, is seen by \cite{Pezzi.etal.16} in their Vlasov simulations, who note that the collision frequency suitable for preventing the ``numerical recurrence of initial conditions" (the echo; see section~\ref{sec:echo} for a detailed discussion of how echoes damp) is a function of the wavenumber of the perturbation. We show that this wavenumber is nothing but $\sim \nu_\rmc/\sigma$, the inverse of the mean free path. Perturbations tend to damp away cubic exponentially within a mean free path, and undergo slow, exponential damping beyond.
\\
\section{Combined linear response}\label{sec:combined_lin_resp}

In the above sections, we separately computed the continuum and discrete mode parts of the linear response. The combined linear response of a plasma is a temporal convolution of the two. This can be readily seen by studying the evolution of the potential perturbation, $\Phi_{1k}(t)$. Note that $\Phi_{1k} = -(2\pi n_e e/\epsilon_0)(G_{1\kappa}(0)/k^2)$, where $G_{1\kappa}(0)$ is the inverse Laplace transform of $\Tilde{G}_{1\kappa}(0)$, given in equation~(\ref{G_1k_0}). To compute the inverse Laplace transform, we make use of the identity,

\begin{align}
{\cal L}^{-1}\left(\Tilde{f}(\gamma)\Tilde{g}(\gamma)\right) &= \frac{1}{2\pi i}\int_{c-i\infty}^{c+i\infty} \rmd\gamma\, \Tilde{f}(\gamma)\Tilde{g}(\gamma) \exp{\left[\gamma t\right]} \nonumber\\
&= \int_0^t \rmd \tau\, f(\tau)\,g(t-\tau),
\label{Laplace_identity}
\end{align}
where $f$ and $g$ are the inverse Laplace transforms of $\Tilde{f}$ and $\Tilde{g}$ respectively. For our problem, $\Tilde{f}$ and $\Tilde{g}$ correspond to $N$ and $1/D$ respectively, with $N$ and $D$ given by equations~(\ref{N_D_0}).

For an initial Gaussian pulse, $h_\kappa(u)$, given by equation~(\ref{initial_hk}), we find that, for $t\gtrsim 1/k\sigma$, the potential response is given by \footnote{$\Phi_{1k}$ is computed by performing the inverse Laplace transforms of $N$ and $1/D$ and using the identity given in equation~[\ref{Laplace_identity}].}

\begin{widetext}
\begin{align}
\Phi_{1k}(t) = -\frac{n_e e}{\epsilon_0} \frac{h_{\kappa 0}}{k^2} \sum_{n} \frac{\exp{\left[\gamma_n t\right]}}{D'\left(\kappa,\gamma_n\right)} \int_0^{k\sigma t} \rmd z\, \exp{\left[-\left(\frac{\gamma_n}{k\sigma}+\kappa\right)z+\frac{\kappa^2}{2}\left(1-e^{-z/\kappa}\right)\left\{\left(1-s^2\right)\left(1-e^{-z/\kappa}\right)+2\right\}-i \kappa u_0 \left(1-e^{-z/\kappa}\right)\right]},
\label{Phi1k_delta}
\end{align}
\end{widetext}
where $\kappa=k\sigma/\nu_\rmc$, $s=\sigma_0/\sigma$, and $u_0 = v_0/\sigma$ with $v_0$ the drift velocity and $\sigma_0$ the velocity dispersion of the initial Gaussian (in $v$) pulse. $\gamma_n$ are the frequencies of the discrete modes that follow the dispersion relation, $D\left(\kappa,\gamma_n\right)=0$, given in equation~(\ref{plasma_disp_rel}). $D'$ refers to the derivative of $D$ with respect to $\gamma_n/k\sigma$ and is given by

\begin{align}
D'\left(\kappa,\gamma_n\right) &= -\frac{\omega^2_\rmP}{k\sigma \nu_\rmc} e^\zeta \left[\zeta^{-a_n}\gamma(a_n,\zeta)\right. \nonumber\\
&\left.- \Gamma^2(a_n)(a_n-\zeta)\, {}_2 F_2 (a_n,a_n;a_n+1,a_n+1;-\zeta)\right],
\end{align}
where $a_n=\gamma_n/\nu_\rmc+\kappa^2$, $\zeta = \kappa^2$, and ${}_2 F_2$ refers to the confluent hypergeometric function of the second kind. 

Let us investigate the temporal behaviour of $\Phi_{1k}$ in some detail. It decays super-exponentially at small time and transitions to an exponential decay at large time. The nature of the initial super-exponential decay, however, depends on $s=\sigma_0/\sigma$, i.e., on how cold the initial perturbation in the DF is relative to the equilibrium DF. Taylor expanding the argument of the exponential, we see that, at $t\lesssim 1/k\sigma$, $\Phi_{1k}(t)$ decays as

\begin{align}
\Phi_{1k}(t) \sim \exp{\left[\,-\frac{k^2\sigma^2 t^2}{2}\left[s^2+\left(\frac{2}{3}-s^2\right)\nu_\rmc t\right]\,\right]}.
\label{Phik_delta_small_time}
\end{align}
For a sufficiently cold perturbation with $s\lesssim \sqrt{2/3}$, the response decays as $\sim \exp{\left[\,-\left(1/3-s^2/2\right)k^2\sigma^2\nu_\rmc t^3\,\right]}$ at small time. Therefore, a perturbation that is Dirac delta in $v$ ($s=0$) initially decays as a cubic exponential, i.e., $\sim \exp{\left[-k^2\sigma^2\nu_\rmc t^3/3\right]}$. If, on the other hand, the initial perturbation is close to the equilibrium Maxwellian form ($s\approx 1$), the response decays as a Gaussian, i.e., $\sim \exp{\left[-k^2\sigma^2 t^2/2\right]}$. This key difference in the initial decay owes to the fact that the potential (also density or temperature) response to a sufficiently cold perturbation decays at small time only due to collisional diffusion, whereas that to a perturbation with a large spread in velocities decays due to phase-mixing as well. The Gaussian decay due to phase-mixing was noted by \cite{Brantov.Bychenkov.06}, \cite{Rozmus.etal.17} and many others, while the cubic exponential decay was noted by \cite{Su.Oberman.68}. We show that these different decay scalings can be obtained from a general form (equations~[\ref{Phi1k_delta}] and [\ref{Phik_delta_small_time}]). Over long time the response to an initial perturbation of any form decays exponentially at the rate of $\min{\left[{\rm Re}\left(\gamma_0\right),{\left(k\sigma\right)}^2/\nu_\rmc\right]}$, under the combined action of collective effects (discrete modes) and collisional damping, where $\gamma_0$ is the frequency of the least damped discrete mode.

Fig.~\ref{fig:Phi1k_vs_t} plots $\Phi_{1k}$ in units of $n_e e h_{\kappa 0}/\epsilon_0 k^2$ as a function of time (in units of $1/k\sigma$) for an initial Dirac delta pulse ($s=0$), for eight different values of $\nu_\rmc/k\sigma$ as indicated and for three different values of $\omega_\rmP/k\sigma$ in three different panels. The initial decay is cubic exponential, which is more pronounced for $\nu_\rmc < k\sigma$. The long time decay is exponential, which manifests more prominently for $\nu_\rmc > k\sigma$. The rate of this long time exponential decay is given by $\min{\left[{\rm Re}\left(\gamma_0\right),{\left(k\sigma\right)}^2/\nu_\rmc\right]}$. For $\nu_\rmc \gtrsim 2\omega_\rmP$, i.e., when the mean free path is somewhat smaller than the Debye length, this decay rate is $\sim {\left(k\sigma\right)}^2/\nu_\rmc$. The transition from the initial cubic exponential decay to the late time exponential decay is more pronounced on scales comparable to the mean free path ($\nu_\rmc \sim k\sigma$). On scales much smaller than the mean free path ($\nu_\rmc \ll k\sigma$), the small time cubic exponential decay is governed by the continuum while the long time exponential decay is governed by the least damped discrete mode (as evident from the temporal oscillations of the long time response). On scales larger than the mean free path ($\nu_\rmc \gg k\sigma$), the response undergoes a slow exponential decay \citep[see also][]{Bell.83,Epperlein.etal.92,Bychenkov.etal.95,Brantov.Bychenkov.06,Brantov.Bychenkov.12,Rozmus.etal.17}, guided by the weakly damped continuum and the least damped discrete mode (the zero-th order non-oscillatory LB mode), over a timescale $\sim \nu_\rmc/{\left(k\sigma\right)}^2$. For $\omega_\rmP\leq \nu_\rmc/2$, the response becomes non-oscillatory, as evident from the absence of oscillations in the response for larger values of $\nu_\rmc$. All in all, Fig.~\ref{fig:Phi1k_vs_t} illustrates the smooth changeover from the super-exponential decay of the response on scales smaller than the mean free path ($\nu_\rmc<k\sigma$) and/or in the diffusive regime ($t<1/\nu_\rmc$), to the exponential decay on scales larger than the mean free path ($\nu_\rmc>k\sigma$) and/or in the friction regime ($t>1/\nu_\rmc$).

\section{Non-linear response: plasma echo}\label{sec:echo}

\begin{figure*}
\centering
\subfloat[$k\sigma T = 1$]{\includegraphics[width=0.95\textwidth]{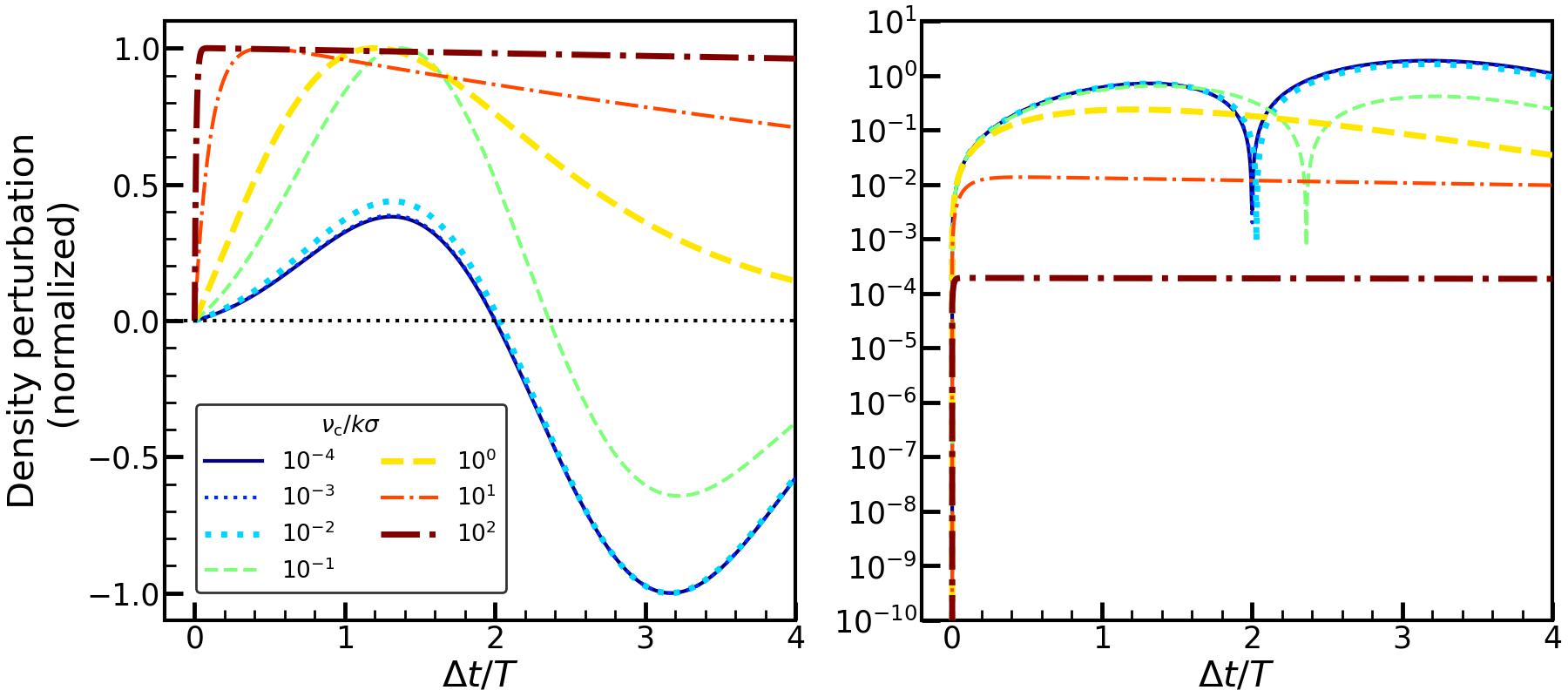}}\\
\subfloat[$k\sigma T = 5$]{\includegraphics[width=0.95\textwidth]{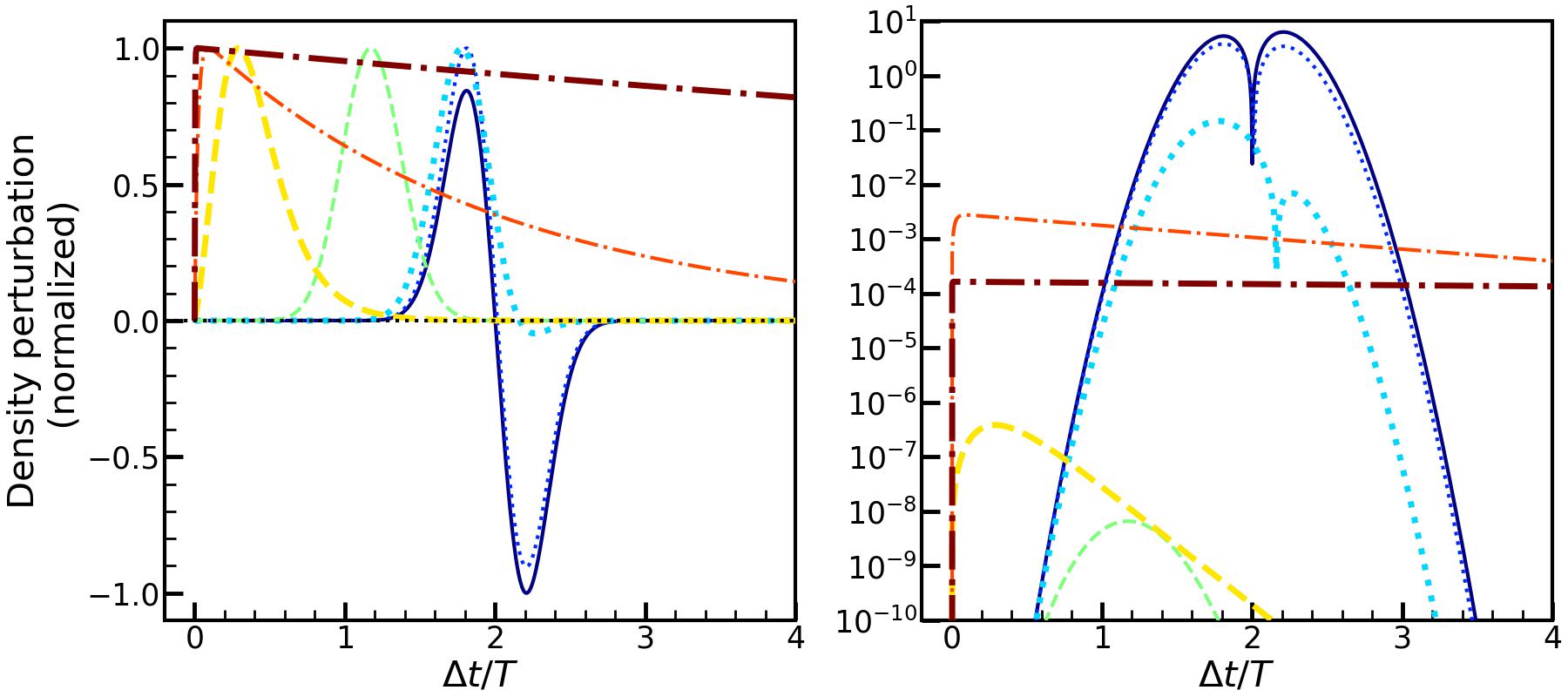}}
\caption{Plasma echo: Second order density perturbation of the $k=\left(k_1-k_2\right)$ mode as a function of $\Delta t/T$, with $\Delta t = t-T$ the time elapsed since the second pulse is introduced at $T$ (see Appendix~\ref{App:second_order_resp} and section~\ref{sec:echo} for details). The first and second pulses have wavenumbers $k_1$ and $k_2$ respectively. Here we have adopted $k_1\sigma/\nu_\rmc = 3$ and $k_2\sigma/\nu_\rmc = 2$. Different lines indicate different values of $\nu_\rmc/k\sigma$. The top and bottom panels correspond to $k\sigma T = 1$ and $5$ respectively. The left panel indicates the density perturbation, $r_{2\kappa}$, normalized by its maximum absolute value, and the right panel indicates $r_{2\kappa}/\left(ab\left(k_1\sigma t_\rmP\right)\left(k_2\sigma t_\rmP\right)\right)$, where $a$ and $b$ are the dimensionless strengths of the first and second potential pulses respectively, and $t_\rmP$ is the temporal pulsewidth that goes to zero, keeping $a t_\rmP$ and $b t_\rmP$ finite. The spikes in the density perturbation indicate the echos. These occur at $\Delta t/T \approx k_2/\left(k_1-k_2\right) = 2$ in the collisionless limit. The echo time approaches $T$ with increasing $\nu_\rmc$. Note that the echoes are detectable even for $\nu_\rmc/k\sigma \sim 1$, as long as $k\sigma T$ is small, i.e., the initial response has not sufficiently decayed via phase-mixing.}
\label{fig:r2k_vs_t}
\end{figure*}

So far, we have discussed how the super-exponential (cubic exponential) and exponential decays with two different timescales show up in the linear response of a weakly collisional plasmas as two asymptotic limits. Now we shall discuss the implications of these two timescales on the non-linear (second order) response. 

The key features of the continuum response in a weakly collisional plasma are: (i) it decays cubic exponentially with time for $\lambda<\lambda_\rmc$ but exponentially for $\lambda>\lambda_\rmc$, and gets damped the most for $\lambda \sim \lambda_\rmc$, and (ii) an externally imposed pulse slows down over time due to the collisional damping of particle velocities over a collision time, $1/\nu_\rmc$. These behaviors not only impact the linear regime of plasma relaxation, but also have implications for the survival and persistence of coherent structures and non-linear phenomena. One such non-linear phenomenon is the plasma echo \cite{Gould.etal.67,Su.Oberman.68,Pavlenko.83,Spentzouris.etal.96}, which attests to the time-reversibility of the Vlasov equation and arises from the continuum response. The fine-grained continuum response to a perturbation gives rise to a macroscopic disturbance, when a second perturbation is introduced into the system sometime after the first perturbation and interferes with the response to it. %This phenomenon, known as the plasma echo, occurs due to a bunching of particles in the phase space.%

Let us analyze the echo problem in some detail, and understand the impact of collisionality on this phenomenon. This requires us to evaluate the second order response of the plasma to an external perturbation. Since we are interested in the echo phenomenon, which is caused by the continuum response, we ignore the self-potentials $\Phi_1$ and $\Phi_2$ in this section\footnote{Including the self-potential would lead to the collective (Landau) damping of the macroscopic density perturbation or the echo, besides the collisional damping we are interested in here.}. We assume that the plasma is excited at $t=0$ by an impulsive kick of sinusoidal electric field with wavenumber $k_1$. The plasma is impulsively kicked for a second time at $t=T$ by another sinusoidal electric field with wavenumber $k_2$. The corresponding $k$ Fourier mode of the potential is given by (see Appendix~\ref{App:second_order_resp} for details)

\begin{align}
\Tilde{\Phi}_k(\gamma) = t_\rmP \left[A\,\delta\left(k-k_1\right)\delta(t) + B\left(k+k_2\right)\delta(t-T)\right],
\end{align}
where $A$ and $B$ are constants, and $t_\rmP$ is the temporal pulsewidth that goes to zero, keeping $A t_\rmP$ and $B t_\rmP$ finite. The resulting second order response, $g_{2\kappa}=\left(\sigma/n_e\right)f_{2k}$ is derived in detail in Appendix~\ref{App:second_order_resp}, and is given by equations~(\ref{g2k_2_app}) and (\ref{Ik_Jkk1_app}). The third term in equation~(\ref{g2k_2_app}) is the relevant term for the plasma echo, and can be expressed as

\begin{widetext}
\begin{align}
&g_{2\kappa}(u,t) = -\frac{a b \left(k_1\sigma t_\rmP\right) \left(k_2\sigma t_\rmP\right)}{\sqrt{2\pi}} \exp{\left[-\frac{u^2}{2}\right]} \; \delta\left(k-\left(k_1-k_2\right)\right) \; \Theta(t-T)\;\calJ_{\kappa,-\kappa_2}(t,T),
\label{g2k_2}
\end{align}
\end{widetext}
where $a$ and $b$ are the dimensionless strengths of the first and second potential pulses respectively (see Appendix~\ref{App:second_order_resp} for details), and $\calJ_{\kappa,-\kappa_2}$ is given by

\begin{widetext}
\begin{align}
&\calJ_{\kappa,-\kappa_2}(t,T) \nonumber\\
&= \Bigg[e^{-\nu_\rmc\left(2t-T\right)}  \nonumber\\
& + \left[iu e^{-\nu_\rmc \left(t-T\right)} + \kappa{\left(1-e^{-\nu_\rmc \left(t-T\right)}\right)}^2 + \kappa_2 e^{-2\nu_\rmc\left(t-T\right)}\left(1-e^{-\nu_\rmc T}\right)\right]\left[iue^{-\nu_\rmc t} + \kappa e^{-\nu_\rmc T}{\left(1-e^{-\nu_\rmc\left(t-T\right)}\right)}^2 - \kappa_2\left(1-e^{-\nu_\rmc T}\right)\left(1-e^{-\nu_\rmc \left(2 t-T\right)}\right)\right]\Bigg]\nonumber\\
&\times \exp{\left[-\kappa^2\left(\nu_\rmc \left(t-T\right) - \frac{\left(1-e^{-\nu_\rmc \left(t-T\right)}\right)\left(3-e^{-\nu_\rmc \left(t-T\right)}\right)}{2}\right)\right]} \exp{\left[-{\kappa}^2_2\left(\nu_\rmc T - \frac{\left(1-e^{-\nu_\rmc T}\right)\left(2+e^{-2\nu_\rmc\left(t-T\right)}\left(1-e^{-\nu_\rmc T}\right)\right)}{2}\right)\right]}\nonumber\\
&\times \exp{\left[\kappa \kappa_2 \left(1-e^{-\nu_\rmc T}\right){\left(1-e^{-\nu_\rmc\left(t-T\right)}\right)}^2\right]}\nonumber\\
&\times \exp{\left[-iu\left(\kappa\left(1-e^{-\nu_\rmc\left(t-T\right)}\right)-\kappa_2 e^{-\nu_\rmc\left(t-T\right)}\left(1-e^{-\nu_\rmc T}\right)\right)\right]}.
\label{Ik_Jkk1}
\end{align}
\end{widetext}
Equations~(\ref{g2k_2}) and (\ref{Ik_Jkk1}) above denote the second order response at the wavenumber, $k=k_1-k_2$, which manifests the echo born out of the interference between the first and second pulses. Information about the initial growth of the echo is carried by the first, second and third factors of equation~(\ref{Ik_Jkk1}). The second and third factors encode information about the damping of the envelope of the echo that occurs over the two timescales discussed in section~\ref{sec:cont_resp}, $\tau_{\rmd,1} \sim {\left(k\sigma\right)}^{-2/3} \nu_\rmc^{-1/3}$ and $\tau_{\rmd,2}\sim \nu_\rmc/{\left(k\sigma\right)}^2$. The fourth factor is a phase-factor that scales as $\sim \exp{\left[-i\frac{v}{\nu_\rmc}\left(\left(k_1-k_2\right)\left(1-e^{-\nu_\rmc\left(t-T\right)}\right)-k_2 e^{-\nu_\rmc\left(t-T\right)}\left(1-e^{-\nu_\rmc T}\right)\right)\right]}$ (note that $k=k_1-k_2$) and becomes independent of $v$ when the argument is zero. This corresponds to a bunching of particles and an enhancement in the macroscopic density perturbation: the echo. It occurs at

\begin{align}
t = t_{\rm echo} \approx T + \frac{1}{\nu_\rmc} \ln\left|1+\frac{k_2}{k_1-k_2}\left(1-e^{-\nu_\rmc T}\right)\right|.
\label{t_echo}
\end{align}
The approximation improves in accuracy for large $\left(k_1-k_2\right)\sigma T$. In the collisionless ($\nu_\rmc T \to 0$) limit, the phase-factor scales as $\sim \exp{\left[-i v\left(\left(k_1-k_2\right) t - k_1 T\right)\right]}$. Accordingly, the echo occurs at $t_{\rm echo}\approx k_1 T /\left(k_1-k_2\right)$, which is also the $\nu_\rmc T \to 0$ limit of equation~(\ref{t_echo}). This is the classic result for echo formation in a collisionless plasma \cite{Gould.etal.67}. In the strong collision ($\nu_\rmc T \to \infty$) limit, $t_{\rm echo} \approx T + \frac{1}{\nu_\rmc}\ln{\left|k_1/\left(k_1-k_2\right)\right|}$. Hence, the echo forms closer to $T$, the time of the second pulse (since the continuum response slows down due to collisional damping), and becomes increasingly difficult to detect with increasing $\nu_\rmc$. 

The echo involves the continuum response becoming independent of $v$, and therefore corresponds to a macroscopic disturbance in the plasma density. Keeping this in mind, the $\left(k_1-k_2\right)$ mode of the second order density perturbation, $r_{2\kappa}=\rho_{2k}/\rho_0$, is computed by integrating equation~(\ref{g2k_2}) over $u$, and is given by equation~(\ref{r2k_app}). The left panels of Fig.~\ref{fig:r2k_vs_t} plot $r_{2\kappa}$, divided by its maximum absolute value, as a function of $\Delta t/T = \left(t-T\right)/T$. The right panels plot the absolute value of $r_{2\kappa}/\left(ab\left(k_1\sigma t_\rmP\right)\left(k_2\sigma t_\rmP\right)\right)$ as a function of $\Delta t/T$. The top and bottom panels correspond to $\left(k_1-k_2\right)\sigma T=1$ and $5$ respectively. We adopt $k_1\sigma/\nu_\rmc = 3$ and $k_2\sigma/\nu_\rmc = 2$. The spikes in the density perturbation are the echoes. In the collisionless limit, the echoes occur at times close to $\Delta t/T \approx k_2 /\left(k_1-k_2\right) = 2$. The time of the echo approaches $T$ with increasing $\nu_\rmc$, which is also evident from equation~(\ref{t_echo}), and becomes indistinguishable from $T$ in the $\nu_\rmc T \to \infty$ limit. The right panels of Fig.~\ref{fig:r2k_vs_t} show that the response gets weaker with increasing $\nu_\rmc$, decaying as $\sim k_1 k_2 \sigma^2/\nu^2_\rmc$ in the $\nu_\rmc T \to \infty$ limit. The echoes become more pulse-like or localized in time for larger $T$. They decay super-exponentially for $\nu_\rmc < \left(k_1-k_2\right)\sigma$ but undergo a slow exponential decay for $\nu_\rmc > \left(k_1-k_2\right)\sigma$. For large $\left(k_1-k_2\right)\sigma T$, the linear response prior to $t=T$, and consequently the echo, are heavily suppressed on scales comparable to the mean free path, i.e., for $\nu_\rmc \sim \left(k_1-k_2\right)\sigma$. This is because the damping timescale is shorter than the time of occurrence of the echo in this case. However, as long as $\left(k_1-k_2\right)\sigma T$ is small, echoes are detectable in a weakly collisional plasma even when $\nu_\rmc/\left(k_1-k_2\right)\sigma \sim 1$, i.e., on scales comparable to the mean free path (see the top panels of Fig.~\ref{fig:r2k_vs_t}). This is because, in this case, the initial response has not sufficiently phase-mixed and damped away due to collisions before the second pulse is introduced.

\section{Discussion and summary}\label{sec:discussion_summary}

We have provided a general formalism for the relaxation of a weakly collisional plasma in the perturbative (linear and second order) regime. Following the seminal work of \cite{Lenard.Bernstein.58}, we compute the linear response of a weakly collisional plasma using the perturbed Boltzmann-Poisson equations and demonstrate that the \cite{Su.Oberman.68} continuum response and the discrete modes \citep[][]{Lenard.Bernstein.58,Ng.Bhattacharjee.Skiff.99,Ng.Bhattacharjee.Skiff.04,Ng.Bhattacharjee.21} are nothing but different limits of a general form of the response. We show that at late times and on scales larger than the Debye length (fluid regime), the continuum response undergoes a much slower decay than what \cite{Su.Oberman.68} predicts. Unlike the \cite{Su.Oberman.68} cubic exponential decay, this decay is a slow exponential just like the discrete modes, and is similar to the rates predicted by \citep[][]{Ng.Bhattacharjee.Skiff.99,Ng.Bhattacharjee.Skiff.04,Ng.Bhattacharjee.21,Bell.83,Epperlein.etal.92,Bychenkov.etal.95,Brantov.Bychenkov.06,Brantov.Bychenkov.12,Rozmus.etal.17}. This finding provides a rigorous justification for the establishment of LTE in the fluid limit. Finally, we compute the second order response of a weakly collisional plasma and demonstrate how the cubic exponential and exponential decays show up in the evolution of the plasma echo. Our findings about the collisional damping of the plasma echo have important implications for dissipation in turbulent plasmas \citep[][]{Schekochihin.etal.16}.

We describe a weakly collisional plasma using a perturbative analysis of the Boltzmann-Poisson equations. We model collisions using a collision operator of the Fokker-Planck type, called the Lenard-Bernstein (LB) operator \citep[][]{Lenard.Bernstein.58}, given by equation~(\ref{LB_coll_op}). The collision frequency, $\nu_\rmc$, parametrizes the rate of collisions. The collision operator consists of two terms: a diffusion term proportional to $\sigma^2\partial^2 f/\partial v^2$ and a damping/friction term proportional to $\partial (vf)/\partial v$. Although one considers both terms to obtain the discrete mode spectrum in the standard theory \citep[][]{Lenard.Bernstein.58}, the analysis of the continuum spectrum and the echo has so far only been performed using the diffusion term \citep[][]{Su.Oberman.68,Short.Simon.02}. This approximation simplifies the calculations but is inconsistent with the fluctuation-dissipation theorem. It is only valid in a limited spatio-temporal domain: at times smaller than the collision time and on scales smaller than the mean free path, i.e., only within the boundary layer. We include both the diffusion and friction terms in the collision operator, and obtain the full response, which is a combination of the continuum response and the discrete modes. In fact, we find that the total response is a temporal convolution of the two. The continuum dominates the initial part of the response, while the long time response has contribution from both the discrete modes and the continuum. 

Collisions affect the discrete modes and the continuum in different ways. The discrete mode response exponentially damps away due to a combination of collective Landau damping and collisional effects; the frequencies and damping rates of the discrete modes get shifted from the Landau values in presence of collisions. When $\nu_\rmc<2\omega_\rmP$, collisions enhance the damping rate of the discrete modes above the Landau damping rate of collisionless plasma. For $\nu_\rmc\gtrsim 2\omega_\rmP$, collisions make the modes non-oscillatory and purely damped. For $\nu_\rmc \gg 2\omega_\rmP$, the damping rates of the modes become weaker than the corresponding Landau damping rates (this slow decay in the frequent collision/fluid regime has also been noted by \citep[][]{Bell.83,Epperlein.etal.92,Bychenkov.etal.95,Brantov.Bychenkov.06,Brantov.Bychenkov.12,Rozmus.etal.17}). Therefore, collisions facilitate the collective damping of the response when they are rare, but oppose damping when they are very frequent.

The continuum part of the response evolves differently with time than the discrete modes. At times smaller than the collision time, $1/\nu_\rmc$, the continuum decays as $\sim \exp{\left[-{\left(t/t_{\rmd,1}\right)}^3\right]}$ over a damping timescale of $t_{\rmd,1} \sim \nu_\rmc^{-1/3}{\left(k\sigma\right)}^{-2/3}$, similar to the prediction by \cite{Su.Oberman.68}. At $t>1/\nu_\rmc$, however, it goes over to an exponential decay over a timescale of $t_{\rmd,2}\sim \nu_\rmc/{\left(k\sigma\right)}^2$. The response damps away the most on scales comparable to the mean free path, i.e.,  for $\nu_\rmc\sim k\sigma$. On smaller scales the decay is super-exponential, as predicted by \cite{Su.Oberman.68}, but on scales larger than the mean free path, the decay is exponential and gets slower with increasing $\nu_\rmc$. We supplement the Fokker-Planck analysis of the continuum response with a complementary analysis of stochastic collisional scattering using the Langevin equation, in the same spirit as the study of Brownian motion \citep[][]{Einstein.1905,Lemons.Gythiel.97}. We model the effect of collisional diffusion by a stochastic acceleration term corresponding to white noise, and that of collisional damping by a friction term. The resulting velocity dispersion increases linearly with time and then saturates to the equilibrium value, $\sigma^2$, after $t\sim 1/\nu_\rmc$. Meanwhile, the dispersion in position increases as $\sim t^3$ for $t\lesssim 1/\nu_\rmc$ and as $\sim t$ for $t\gtrsim 1/\nu_\rmc$. This is a classic feature of random walk or diffusion in the velocity space, followed by damping or friction in the velocity space and diffusion in the position space. The cubic exponential decay of the continuum response for $t<1/\nu_\rmc$, followed by an exponential decay for $t>1/\nu_\rmc$, is a consequence of collisional diffusion followed by damping. We emphasize that the temporal nature of continuum decay obtained in this paper is not particularly sensitive to the Lenard-Bernstein form of the collision operator, but is a robust phenomenon (at least qualitatively) in any system governed by short range random/chaotic collisions.

The general treatment of plasma relaxation discussed in this paper shows that the \cite{Su.Oberman.68} scaling of continuum decay is only valid for small time and on scales smaller than the mean free path. This is expected since the \cite{Su.Oberman.68} calculation neglects the damping term in the LB operator, and entirely relies on a boundary-layer analysis, as also demonstrated by \cite{Short.Simon.02}. Therefore, one should be cautious about the limited range of validity of the \cite{Su.Oberman.68} result when addressing phenomena pertaining to collisional diffusion in plasmas. One such phenomenon is the non-linear effect of plasma echo, a macroscopic disturbance that occurs due to the interference of the continuum responses to two subsequent pulses of electric field with wavenumber $k_1$ and $k_2$ and has a wavenumber $k=k_1-k_2$. On scales smaller than the Debye length, i.e., for small $\nu_\rmc$ ($\nu_\rmc < k\sigma$), the echo damps cubic exponentially as \cite{Su.Oberman.68} predicts, but on larger scales and/or large $\nu_\rmc$ ($\nu_\rmc > k\sigma$), it damps far slower. However, with increasing $\nu_\rmc$, the echo time approaches the time of initiation of the second pulse (since the continuum response slows down due to collisional damping), and becomes harder, albeit possible, to detect. Contrary to what \cite{Su.Oberman.68} predict, we infer that the echo is detectable even for $\nu_\rmc$ as large as $\sim \left(k_1-k_2\right)\sigma$, as long as the second pulse is introduced within a few phase-mixing timescales after the first.

The essential features of plasma relaxation described in this paper can be understood by comparing the plasma frequency, $\omega_\rmP$, the collision frequency, $\nu_\rmc$, and the phase-mixing frequency, $k\sigma$, or alternatively by comparing three different length scales: the Debye length, $\lambda_\rmD = \sigma/\omega_\rmP$, the mean free path, $\lambda_\rmc = \sigma/\nu_\rmc$, and the perturbation scale, $\lambda \sim 1/k$. The self-limiting nature of collisions is manifest in the stagnated damping of both the discrete modes and the continuum beyond certain scales. The damping of the discrete modes occurs efficiently for $\nu_\rmc\lesssim 2\omega_\rmP$, but gets stagnated for $\nu_\rmc\gtrsim 2\omega_\rmP$. In other words, collisions facilitate the collective Landau-like damping of the discrete modes when the mean free path, $\lambda_\rmc = \sigma/\nu_\rmc$, is larger than half the Debye length, $\lambda_\rmD = \sigma/\omega_\rmP$, and oppose collective damping when $\lambda_\rmc\lesssim \lambda_\rmD/2$. The discrete modes, however, become non-oscillating for $\lambda_\rmc\lesssim \lambda_\rmD/2$, since multiple collisional scatterings within the Debye scale destroy the coherence of the collective plasma oscillations. The collisional damping of the continuum response occurs rapidly for $\nu_\rmc\sim k\sigma$ or $\lambda \sim \lambda_\rmc$, but is arrested for $\nu_\rmc>k\sigma$ or $\lambda>\lambda_\rmc$. This transition of the damping rate at $\lambda \sim \lambda_\rmc$ owes to the fact that the collisional diffusion of particle velocities is restricted within the mean free path. This means that an LTE is established only within the mean free path since the DF approaches the Maxwellian form locally (within $\lambda_\rmc$ around a point), after a collision time, $1/\nu_\rmc$. As a result, a local equation of state can be adopted and a fluid formalism implemented to study the evolution of plasma perturbations on scales above the mean free path. An ideal fluid description is, however, only valid as long as $\nu_\rmc\ll 2\omega_\rmP$, or the mean free path, $\lambda_\rmc$, is substantially larger than $\lambda_\rmD/2$. When $\nu_\rmc \gtrsim 2\omega_\rmP$, or $\lambda_\rmc \lesssim \lambda_\rmD/2$, the discrete modes are rendered non-oscillatory by collisions. Therefore, an ideal fluid description becomes questionable in this regime. To properly study the long time relaxation of plasmas in this regime, one should either use a collisional kinetic theory formalism as described in this paper \citep[see also][]{Bell.83,Epperlein.etal.92,Bychenkov.etal.95,Brantov.Bychenkov.06,Brantov.Bychenkov.12,Rozmus.etal.17}, or incorporate viscous dissipation in the corresponding fluid formalism. 

We have developed a general formalism for the perturbative (linear and second order) relaxation of weakly collisional plasmas in this paper. It would be interesting to investigate the effect of weak collisions also on the non-linear relaxation of plasmas. The crucial role of collisional damping or friction in breaking the symmetry of wave-particle resonances (shifting and splitting of resonance lines) during the quasilinear relaxation of plasmas has been noted by \cite{Duarte.etal.23}. Their analysis is, however, restricted to the domain where the collision frequency significantly exceeds the libration/bounce frequency of charged particles in the trapped region of the phase space. In future work, we intend to investigate quasilinear relaxation and non-linear relaxation/particle trapping when these two frequencies are comparable. In particular, we would like to address the following questions. How does weak collisionality affect the trapping of ions and electrons in the non-linear Bernstein-Green-Kruskal or BGK \cite{Bernstein.etal.57} modes? Is the BGK mode stable in presence of collisions? Do collisions untrap the otherwise trapped particles, and if so, how fast? How do collisions impact the so-called sideband instability \cite{Wharton.etal.68} of BGK modes? We envision developing a theory for phase space cascade and turbulence in a weakly collisional plasma, similar to that advocated by \cite{Nastac.etal.23}, but relaxing some of their assumptions. We plan to extend the formalism presented in this paper to the study of the impact of weak collisions on self-gravitating systems, since such systems are governed by the Boltzmann-Poisson equations just like plasmas (with the crucial difference that the right-hand-side of the Poisson equation carries the opposite sign relative to the plasma case). This formalism can be applied to the evolution of various astrophysical systems such as globular clusters and self-interacting dark matter halos, where collisions play an important role.
\\
\begin{acknowledgments}
We wish to acknowledge Scott Tremaine, Chris Hamilton, Vin{\'i}cius Duarte, Martin Weinberg and Wrick Sengupta for insightful discussions and valuable suggestions. This research is supported by the National Science Foundation Award 2209471 and Princeton University.
\end{acknowledgments}

\appendix

\section{Calculation of the linear response}\label{App:lin_resp_calc}

The linear response of a weakly collisional plasma is computed by simultaneously solving the linearized Boltzmann and Poisson equations, which amounts to solving equation~(\ref{master_eq_nd}) for $\Tilde{g}_{1\kappa}(u)$. Upon taking the Fourier transform in $u$, this becomes equation~(\ref{master_eq_nd_w}), which needs to be solved for $\Tilde{G}_{1\kappa}(w)$, the Fourier transform of $\Tilde{g}_{1\kappa}(u)$. The solution for $\Tilde{G}_{1\kappa}(w)$ is given by equation~(\ref{G_1k_w}), which involves $\Tilde{G}_{1\kappa}(0)$. Therefore, substituting $w=0$ in equation~(\ref{G_1k_w}), we obtain the expression for $\Tilde{G}_{1\kappa}(0)$ given in equations~(\ref{G_1k_0}) and (\ref{N_D_0}). Re-substituting the expression for $\Tilde{G}_{1\kappa}(0)$ in equation~(\ref{G_1k_w}) now yields the following expression for $\Tilde{G}_{1\kappa}(w)$:

\begin{align}
\Tilde{G}_{1\kappa}(w) &= \Tilde{G}^{(1)}_{1\kappa}(w) + \Tilde{G}^{(2)}_{1\kappa}(w) + \Tilde{G}^{(3)}_{1\kappa}(w),
\label{G_1k_0_app}
\end{align}
where

\begin{widetext}
\begin{align}
\Tilde{G}_{1\kappa}^{(1)}(w) &= \exp{\left[-\frac{w^2}{2} - \kappa w\right]}\;\; {\left(w-\kappa\right)}^{-\left(\eta+\kappa^2\right)} \int_\kappa^w \rmd w'\, \exp{\left[\frac{w'^2}{2} + \kappa w'\right]}\;\; {\left(w'-\kappa\right)}^{\,\eta+\kappa^2-1} H_\kappa(w'),\nonumber\\
\Tilde{G}_{1\kappa}^{(2)}(w) &= -\frac{2\pi}{\kappa}{\left(\frac{\omega_\rmP}{\nu_\rmc}\right)}^2 \frac{N^{(2)}(\kappa,\eta)}{D(k,\eta)} \times \exp{\left[-\frac{w^2}{2} - \kappa w\right]}\;\; {\left(w-\kappa\right)}^{-\left(\eta+\kappa^2\right)} \int_\kappa^w \rmd w'\, \exp{\left[\frac{w'^2}{2} + \kappa w'\right]}\;\; {\left(w'-\kappa\right)}^{\,\eta+\kappa^2-1} w' G_0(w'),\nonumber\\
\Tilde{G}_{1\kappa}^{(3)}(w) &= -\frac{\kappa A_\kappa(\eta)}{D(k,\eta)} \times \exp{\left[-\frac{w^2}{2} - \kappa w\right]}\;\; {\left(w-\kappa\right)}^{-\left(\eta+\kappa^2\right)} \int_\kappa^w \rmd w'\, \exp{\left[\frac{w'^2}{2} + \kappa w'\right]}\;\; {\left(w'-\kappa\right)}^{\,\eta+\kappa^2-1} w' G_0(w').
\label{G_1k_w_f_app}
\end{align}
\end{widetext}
Here,

\begin{align}
N^{(2)}(\kappa,\eta) &= {\left(-k\right)}^{-\left(\eta+\kappa^2\right)} \nonumber\\
&\times \int_\kappa^0 \rmd w'\, \exp{\left[\frac{w'^2}{2} + \kappa w'\right]}\;\; {\left(w'-\kappa\right)}^{\,\eta+\kappa^2-1} H_\kappa(w'),
\end{align}
and

\begin{align}
D(\kappa,\eta) &= 1 + \frac{2\pi}{\kappa}{\left(\frac{\omega_\rmP}{\nu_\rmc}\right)}^2 {\left(-\kappa\right)}^{-\left(\eta+\kappa^2\right)} \nonumber\\
&\times \int_\kappa^0 \rmd w'\, \exp{\left[\frac{w'^2}{2} + \kappa w'\right]}\;\; {\left(w'-\kappa\right)}^{\,\eta+\kappa^2-1} w' G_0(w') \nonumber\\
&= 1+ {\left(\frac{\omega_\rmP}{\kappa\nu_\rmc}\right)}^2 \left[\,1-\left(a-\zeta\right)\zeta^{-a}\exp{\left[\zeta\right]}\,\gamma\left(a,\zeta\right)\,\right],
\label{N_D_0_app}
\end{align}
where $a=\eta+\kappa^2$, $\zeta = \kappa^2$, and $\gamma(a,\zeta)=\int_0^\zeta \rmd z\,z^{a-1}\exp{\left[-z\right]}$ is the lower incomplete gamma function. We have assumed that the unperturbed DF, $g_0$, is of Maxwellian form, i.e., $g_0(u) = \exp{\left[-u^2/2\right]}\Big/\sqrt{2\pi}$, whose velocity Fourier transform is given by

\begin{align}
G_0(w) = \frac{1}{2\pi}\int_{-\infty}^\infty \rmd u\, \exp{\left[-iwu\right]}\, g_0(u) = \frac{1}{2\pi} \exp{\left[-w^2/2\right]}.
\end{align}

The temporal response can be computed by taking the inverse Laplace transform and inverse Fourier transform of $\Tilde{G}_{1\kappa}(w)$. The three different terms in the expression for $\Tilde{G}_{1\kappa}(w)$ given in equation~(\ref{G_1k_0_app}) then respectively correspond to three different parts of the response: $\Tilde{G}_{1\kappa}^{(1)}$ corresponds to the continuum response, $\Tilde{G}_{1\kappa}^{(2)}$ corresponds to the discrete modes temporally convoluted with the continuum, and $\Tilde{G}_{1\kappa}^{(3)}$ corresponds to the direct response to the perturber or `wake', temporally convoluted with the continuum and the discrete modes.

Computation of the continuum part of the response amounts to taking the inverse Laplace transform and inverse Fourier transform (in $u$) of $\Tilde{G}_{1\kappa}^{(1)}$ in equation~(\ref{G_1k_w_f_app}). The inverse Laplace transform of ${\left(w-k\right)}^{-\left(\eta+\kappa^2\right)} {\left(w'-k\right)}^{\left(\eta+\kappa^2-1\right)}$ (recall that $\gamma=\nu_\rmc\eta$) yields

\begin{align}
&{\cal L}^{-1}\left[{\left(w-\kappa\right)}^{-\left(\eta+\kappa^2\right)} {\left(w'-\kappa\right)}^{\left(\eta+\kappa^2-1\right)}\right] \nonumber\\
&= \frac{1}{2\pi i}\int_{c-i\infty}^{c+i\infty} \rmd\eta\, {\left(w-\kappa\right)}^{-\left(\eta+\kappa^2\right)} {\left(w'-\kappa\right)}^{\left(\eta+\kappa^2-1\right)} \exp{\left[\eta \nu_\rmc t\right]} \nonumber\\
&= \exp{\left[-\kappa^2 \nu_\rmc t\right]}\; \delta\left[w'-\left(we^{-\nu_\rmc t}+\kappa\left(1-e^{-\nu_\rmc t}\right)\right)\right].
\label{inv_Lap_id_app}
\end{align}
Next, we integrate the inverse Laplace transform of $\Tilde{G}_{1\kappa}^{(1)}$ over $w'$ and take the inverse Fourier transform in $u$ of the result to obtain the following expression for the continuum response:

\begin{widetext}
\begin{align}
g_{1\kappa}^{\rm cont}(u,t) &= \int_{-\infty}^\infty \rmd w\,\exp{\left[iwu\right]} \frac{1}{2\pi i}\int_{c-i\infty}^{c+i\infty} \rmd\eta\, \exp{\left[\eta \nu_\rmc t\right]} \, \Tilde{G}_{1\kappa}^{(1)}(w,\eta) \\
&= \exp{\left[-\kappa^2 \nu_\rmc t\right]}\int_{-\infty}^\infty \rmd w\,\exp{\left[iwu\right]} \exp{\left[-\frac{w^2}{2} - \kappa w\right]} \int_\kappa^w \rmd w'\exp{\left[\frac{w'^2}{2} + \kappa w'\right]} \delta\left[w'-\left(we^{-\nu_\rmc t}+\kappa\left(1-e^{-\nu_\rmc t}\right)\right)\right]\,H_\kappa(w') \nonumber\\ \nonumber\\
&= \sqrt{\frac{2}{1-e^{-2\nu_\rmc t}}} \exp{\left[-\frac{u^2}{2\left(1-e^{-2\nu_\rmc t}\right)}\right]} \exp{\left[-\kappa^2\left(\nu_\rmc t - 2\tanh{\frac{\nu_\rmc t}{2}}\right)\right]} \exp{\left[-i \kappa u\, \tanh{\frac{\nu_\rmc t}{2}}\right]}\, \calF(\kappa,\nu_\rmc t),
\label{g1k_cont_app}
\end{align}
\end{widetext}
where

\begin{align}
&\calF(w,\kappa,\nu_\rmc t) \nonumber\\
&= \int_{-\infty}^\infty \rmd z \, \exp{\left[-{\left\{z+\left(\sqrt{2\tanh{\frac{\nu_\rmc t}{2}}}\kappa - i\frac{u}{\sqrt{2\left(1-e^{-2\nu_\rmc t}\right)}}\right)\right\}}^2\right]}\nonumber\\
&\times H_\kappa\left(\kappa+ze^{-\nu_\rmc t}\right).
\end{align}
In going from the second to third line of equation~(\ref{g1k_cont_app}), we have made the substitution, $z=w-\kappa$. The continuum response is evaluated in section~\ref{sec:init_pulse_models} for different models of $h_k(u)$, whose velocity Fourier transform is $H_k(w)$.

\section{Calculation of velocity and position dispersions in the Langevin formalism} \label{App:vel_pos_disp_Langevin_calc}

In section~\ref{sec:Langevin_analysis}, we gain physical insight into the workings of collisional damping and diffusion by studying the evolution of the first and second moments of the continuum DF response, rather than that of the continuum response itself, which we study in section~\ref{sec:init_pulse_models}. To achieve this, we solve the stochastic Langevin equation, given in equation~(\ref{Langevin_eq}), assuming a collisional friction proportional to $-v$ and a white noise power spectrum for the stochastic acceleration due to collisions. Here we solve this equation to derive the velocity dispersion, $\sigma^2(t)$, and the position dispersion, $\sigma^2_x(t)$.

The Langevin equation~(\ref{Langevin_eq}) can be integrated once with respect to time to yield the following expression for the velocity of a particle:

\begin{align}
v(t) = \frac{\rmd x}{\rmd t} = v_0\,e^{-\nu_\rmc t} + \int_0^t \rmd t'\,e^{-\nu_\rmc(t-t')} \xi(t').
\label{v_Langevin_app}
\end{align}
The velocity dispersion, $\sigma(t)$, can now be computed as follows:

\begin{align}
\sigma^2(t) &= \left<\left(v(t)-v_0\,e^{-\nu_\rmc t}\right)^2\right> \nonumber\\
&= \int_0^t \rmd t'\,e^{-\nu_\rmc(t-t')} \int_0^t \rmd t''\,e^{-\nu_\rmc(t-t'')} \left<\xi(t')\,\xi(t'')\right> \nonumber\\
&= \calD \int_0^t \rmd t'\,e^{-\nu_\rmc(t-t')} \int_0^t \rmd t''\,e^{-\nu_\rmc(t-t'')} \delta\left(t'-t''\right) \nonumber\\
&= \calD \int_0^t \rmd t' e^{-2\nu_\rmc t'}\nonumber\\
&= \sigma^2 \left(1-e^{-2\nu_\rmc t}\right).
\label{vel_disp_Langevin_app}
\end{align}
In going from the second to the third line, we have used the assumption of a white noise power spectrum for $\xi(t)$, i.e., $\left<\xi(t)\xi(t')\right>=\calD\delta(t-t')$, and in going from the fourth to the fifth line, we have used the Einstein relation that $\calD = \nu_\rmc \sigma^2$.

Equation~(\ref{v_Langevin_app}) can be integrated one more time with respect to time to yield the position of a particle:

\begin{align}
x(t) &= \frac{v_0}{\nu_\rmc}\left(1-e^{-\nu_\rmc t}\right) + \int_0^t \rmd t'\int_0^{t'} \rmd t''\,e^{-\nu_\rmc\left(t'-t''\right)} \xi(t'').
\label{x_Langevin_app}
\end{align}
The dispersion in the position, $\sigma_x(t)$, is computed as follows:

\begin{widetext}
\begin{align}
\sigma^2_x(t) &= \left<{\left[x(t)-\frac{v_0}{\nu_\rmc}\left(1-e^{-\nu_\rmc t}\right)\right]}^2\right> \nonumber\\
&= \int_0^t \rmd t_1 \int_0^{t_1} \rmd t_2 \,e^{-\nu_\rmc\left(t_1-t_2\right)} \int_0^t \rmd t_3 \int_0^{t_3} \rmd t_4 \,e^{-\nu_\rmc\left(t_3-t_4\right)} \left<\xi(t_2)\xi(t_4)\right>\nonumber\\
&= \calD\int_0^t \rmd t_1 \int_0^{t_1} \rmd t_2 \,e^{-\nu_\rmc\left(t_1-t_2\right)} \int_0^t \rmd t_3 \int_0^{t_3} \rmd t_4 \,e^{-\nu_\rmc\left(t_3-t_4\right)}\, \delta\left(t_4-t_2\right),
\label{mean_sq_x_Langevin_1_app}
\end{align}    
\end{widetext}
where, in going from the second to the third line, we have used the assumption that $\left<\xi(t)\xi(t')\right>=\calD\delta(t-t')$. Noting that the integral over $t_4$ is non-zero only when $t_2 \leq t_3$, and using the Einstein relation, $\calD = \nu_\rmc \sigma^2$, we perform the integral over $t_4$ to obtain

\begin{align}
\sigma^2_x(t) &= \nu_\rmc \sigma^2 \int_0^t \rmd t_1 \, e^{-\nu_\rmc t_1} \int_0^{t_1} \rmd t_2 \, e^{2\nu_\rmc t_2} \int_{t_2}^t \rmd t_3 \, e^{-\nu_\rmc t_3}.
\end{align}
Performing the integral over $t_3$ yields

\begin{align}
\sigma^2_x(t) &= \sigma^2 \int_0^t \rmd t_1 \, e^{-\nu_\rmc t_1} \int_0^{t_1} \rmd t_2 \, \left(e^{\nu_\rmc t_2} - e^{\nu_\rmc\left(2 t_2 - t\right)}\right),
\end{align}
and performing the $t_2$ integral yields

\begin{align}
\sigma^2_x(t) = \frac{\sigma^2}{\nu_\rmc} \int_0^t \rmd t_1\, \left[1-\frac{3}{2}e^{-\nu_\rmc t_1} + \frac{1}{2} e^{-\nu_\rmc \left(t+t_1\right)}\right].
\end{align}
Finally, performing the above $t_1$ integral yields the following expression for $\sigma^2_x(t)$:

\begin{align}
\sigma^2_x(t) &= {\left(\frac{\sigma}{\nu_\rmc}\right)}^2 \left[\nu_\rmc t - \frac{\left(1-e^{-\nu_\rmc t}\right)\left(3-e^{-\nu_\rmc t}\right)}{2}\right].
\label{mean_sq_x_Langevin_app}
\end{align}

\section{Calculation of the second order response and the plasma echo}\label{App:second_order_resp}

The linear response of a plasma consists of a free streaming continuum and damped discrete modes. When marginalized over the velocities, the continuum response phase-mixes away. The non-linear response is, however, fundamentally different in that the continuum response manifests a bunching of particles in the phase space at specific times, leading to macroscopic disturbances in the density. This phenomenon is known as the plasma echo, and can be readily seen at second order in perturbation. 

The evolution of the second order response, $f_2$, is governed by the following equation:

\begin{align}
\frac{\partial f_2}{\partial t} + v\frac{\partial f_2}{\partial x}  = -\frac{e}{m}\frac{\partial f_1}{\partial v} \frac{\partial \Phi_\rmP}{\partial x} + \nu_\rmc \frac{\partial}{\partial v}\left(v f_2 + \sigma^2 \frac{\partial f_2}{\partial v}\right).
\label{lin_CBE_act_ang_app}
\end{align}
Since the echo phenomenon is caused by the continuum response, we ignore the self-potential $\Phi_1$ of the plasma that leads to discrete modes. Non-dimensionalizing the above equation, then taking the Fourier transform in $x$ and $u=v/\sigma$ and the Laplace transform in $t$ as detailed in section~\ref{sec:lin_eq_solve}, we obtain the following differential equation for $\Tilde{G}_{2k}(w)$, the Fourier-Laplace mode of $f_2$:

\begin{align}
&(w-\kappa)\,\frac{\partial \Tilde{G}_{2\kappa}}{\partial w} + (w^2+\eta)\,\Tilde{G}_{2\kappa}(w,\eta) \nonumber\\
&= - w \int \rmd \kappa'\, \left(\kappa-\kappa'\right) \int \rmd \eta' A_{\kappa-\kappa'}(\eta-\eta')\,\Tilde{G}_{1\kappa'}(w,\eta').
\label{master_eq_nd_w_app}
\end{align}
Here, we have assumed that $f_2(t=0)=0$. The above differential equation can be integrated using the method of integrating factor to yield the following solution for $\Tilde{G}_{2\kappa}$:

\begin{widetext}
\begin{align}
\Tilde{G}_{2\kappa}(w,\eta) &= -\exp{\left[-\frac{w^2}{2} - \kappa w\right]}\;\; {\left(w-\kappa\right)}^{-\left(\eta+\kappa^2\right)} \int_\kappa^w \rmd w'\, \exp{\left[\frac{w'^2}{2} + \kappa w'\right]}\;\; {\left(w'-\kappa\right)}^{\,\eta+\kappa^2-1} w' \nonumber\\
&\times \int \rmd \kappa'\,\left(\kappa-\kappa'\right) \int \rmd \eta' A_{\kappa-\kappa'}(\eta-\eta')\, \Tilde{G}_{1\kappa'}(w',\eta').
\label{G_2k_eta_app}
\end{align}
\end{widetext}
The solution for $\Tilde{G}_{1\kappa}$ is given by

\begin{widetext}
\begin{align}
\Tilde{G}_{1\kappa}(w,\eta) &= -\kappa A_\kappa(\eta)\, \exp{\left[-\frac{w^2}{2} - \kappa w\right]}\;\; {\left(w-\kappa\right)}^{-\left(\eta+\kappa^2\right)} \int_\kappa^w \rmd w'\, \exp{\left[\frac{w'^2}{2} + \kappa w'\right]}\;\; {\left(w'-\kappa\right)}^{\,\eta+\kappa^2-1} w' G_0(w'),
\end{align}
\end{widetext}
where we have assumed that $H_\kappa(w)=0$, i.e., $f_1(t=0)=0$.

For simplicity, we assume hereon that the perturbing potential is impulsive in time. Let the perturbing potential be a combination of two impulses, one at $t=0$ and the other at $t=T$. This implies that

\begin{align}
\Phi_k(t) = t_\rmP \left[N_k^{(1)}\delta(t) + N_k^{(2)}\delta(t-T)\right].
\end{align}
Here, $N_k^{(1)}$ and $N_k^{(2)}$ are the spatial Fourier transforms of the first and second pulses respectively, and $t_\rmP$ is the temporal width of each pulse, which tends towards zero, keeping $N_k^{(1)} t_\rmP$ and $N_k^{(2)} t_\rmP$ finite. The Laplace transform of this, for $t>T$, is given by

\begin{align}
\Tilde{\Phi}_k(\gamma) = t_\rmP \left[N_k^{(1)}+N_k^{(2)}e^{-\gamma T}\Theta(t-T)\right].
\end{align}
In non-dimensional form, we have

\begin{align}
A_\kappa(\eta) = \frac{\nu_\rmc e}{m\sigma^2} \Tilde{\Phi}_k(\gamma) = \nu_\rmc t_\rmP \left[\calA_\kappa + \calB_\kappa e^{-\eta \nu_\rmc T}\Theta(t-T)\right],
\end{align}
where $\calA_\kappa = (e/m\sigma^2) N_k^{(1)}$ and $\calB_\kappa = (e/m\sigma^2) N_k^{(2)}$.

Let us first evaluate the linear response, $G_{1\kappa}(w,t)$, which is an essential ingredient of the second order response. Using the result for inverse Laplace transform from equation~(\ref{inv_Lap_id_app}), we get that

\begin{align}
G_{1\kappa}(w,t) &= -\kappa \nu_\rmc t_\rmP \exp{\left[-\frac{w^2}{2}\right]} \nonumber\\
&\times \left[\calA_\kappa \calL_
\kappa(w,t) + \calB_\kappa \calL_\kappa(w,t-T)\,\Theta(t-T)\right],
\end{align}
where 
\begin{align}
\calL_\kappa(w,t) &= \left(w e^{-\nu_\rmc t} + \kappa\left(1-e^{-\nu_\rmc t}\right)\right)\nonumber\\
&\times \exp{\left[-\kappa \left(w-\kappa\right)\left(1-e^{-\nu_\rmc t}\right) - \kappa^2 \nu_\rmc t \right]}.
\end{align}

Now, let us evaluate the second order response. We inverse Laplace transform equation~(\ref{G_2k_eta_app}) to obtain

\begin{widetext}
\begin{align}
G_{2\kappa}(w,t) &= -\nu_\rmc t_\rmP \exp{\left[-\frac{w^2}{2} - \kappa w\right]} \int_\kappa^w \rmd w'\, \exp{\left[\frac{w'^2}{2} + \kappa w'\right]}\, w' \int \rmd \kappa' \left(\kappa-\kappa'\right) \nonumber\\
&\times \left[\calA_{\kappa-\kappa'} \exp{\left[-\kappa^2 \nu_\rmc t\right]}\, \delta\left[w'-\left(w e^{-\nu_\rmc t}+\kappa\left(1-e^{-\nu_\rmc t}\right)\right)\right]G_{1\kappa'}(w',0)\right.\nonumber\\
&+\left.\Theta\left(t-T\right) \calB_{\kappa-\kappa'}\exp{\left[-\kappa^2 \nu_\rmc \left(t-T\right)\right]}\, \delta\left[w'-\left(w e^{-\nu_\rmc \left(t-T\right)}+\kappa\left(1-e^{-\nu_\rmc \left(t-T\right)}\right)\right)\right]G_{1\kappa'}(w',T)\right].
\end{align}
\end{widetext}
Performing the $w'$ integral and the velocity Fourier transform of the above, we obtain the second order response as follows:

\begin{widetext}
\begin{align}
g_{2\kappa}(u,t) &= \frac{1}{\sqrt{2\pi}} \exp{\left[-\frac{u^2}{2}\right]}\, {\left(\nu_\rmc t_\rmP\right)}^2 \int \rmd\kappa'\kappa'\left(\kappa-\kappa'\right) \Bigg[\,\calA_{\kappa-\kappa'}\calA_{\kappa'}\calI_{\kappa}(t) + \Theta\left(t-T\right)\calB_{\kappa-\kappa'}\Big(\calB_{\kappa'}\calI_{\kappa}(t-T)+\calA_{\kappa'}\calJ_{\kappa, \kappa'}(t,T)\Big)\,\Bigg],
\label{g2k_1_app}
\end{align}
\end{widetext}
where 

\begin{widetext}
\begin{align}
&\calI_{\kappa}(t) = \left[ e^{-2\nu_\rmc t} + {\left(iu e^{-\nu_\rmc t} + \kappa\,{\left(1-e^{-\nu_\rmc t}\right)}^2\right)}^2\right] \exp{\left[-\kappa^2\left(\nu_\rmc t - \frac{\left(1-e^{-\nu_\rmc t}\right)\left(3-e^{-\nu_\rmc t}\right)}{2}\right)\right]} \exp{\left[-i\kappa u \left(1-e^{-\nu_\rmc t}\right)\right]},\nonumber\\ \nonumber\\
&\calJ_{\kappa,\kappa'}(t,T) \nonumber\\
&= \left[e^{-\nu_\rmc\left(2t-T\right)} \right. \nonumber\\
& \left. + \left[iu e^{-\nu_\rmc \left(t-T\right)} + \kappa{\left(1-e^{-\nu_\rmc \left(t-T\right)}\right)}^2 - \kappa'e^{-2\nu_\rmc\left(t-T\right)}\left(1-e^{-\nu_\rmc T}\right)\right]\left[iue^{-\nu_\rmc t} + \kappa e^{-\nu_\rmc T}{\left(1-e^{-\nu_\rmc\left(t-T\right)}\right)}^2+\kappa'\left(1-e^{-\nu_\rmc T}\right)\left(1-e^{-\nu_\rmc \left(2 t-T\right)}\right)\right]\right]\nonumber\\
&\times \exp{\left[-\kappa^2\left(\nu_\rmc \left(t-T\right) - \frac{\left(1-e^{-\nu_\rmc \left(t-T\right)}\right)\left(3-e^{-\nu_\rmc \left(t-T\right)}\right)}{2}\right)\right]} \exp{\left[-{\kappa'}^2\left(\nu_\rmc T - \frac{\left(1-e^{-\nu_\rmc T}\right)\left(2+e^{-2\nu_\rmc\left(t-T\right)}\left(1-e^{-\nu_\rmc T}\right)\right)}{2}\right)\right]}\nonumber\\
&\times \exp{\left[-\kappa \kappa' \left(1-e^{-\nu_\rmc T}\right){\left(1-e^{-\nu_\rmc\left(t-T\right)}\right)}^2\right]}\nonumber\\
&\times \exp{\left[-iu\left(\kappa\left(1-e^{-\nu_\rmc\left(t-T\right)}\right)+\kappa'e^{-\nu_\rmc\left(t-T\right)}\left(1-e^{-\nu_\rmc T}\right)\right)\right]}.
\label{Ik_Jkk1_app}
\end{align}
\end{widetext}
Note that $\calI_\kappa(t)=\calJ_{\kappa,\kappa'=0}(t,T=0)$. 

The above response is a convolution over $k$, and can be simplified if we adopt a perturber potential that is sinusoidal in $x$. Assuming that $\Phi_\rmP\sim \exp{\left[i k_1 x\right]}$ at $t=0$ and $\Phi_\rmP \sim \exp{\left[-i k_2 x\right]}$ at time $T$, we have that $\calA_\kappa = a\, \delta(k-k_1)$ and $\calB_\kappa = b\, \delta(k+k_2)$. Substituting these in equation~(\ref{g2k_1_app}) and performing the integration over $\kappa'$ yields the following form for the second order response:

\begin{align}
&g_{2\kappa}(u,t) = \frac{1}{\sqrt{2\pi}} \exp{\left[-\frac{u^2}{2}\right]}\nonumber\\
&\times \Bigg[ a^2{\left(k_1\sigma t_\rmP\right)}^2 \delta\left(k-2 k_1\right) \calI_\kappa(t)\nonumber\\
&+ b^2 {\left(k_2\sigma t_\rmP\right)}^2\delta\left(k+2 k_2\right) \Theta(t-T)\, \calI_\kappa(t-T) \nonumber\\
&- a b \left(k_1\sigma t_\rmP\right) \left(k_2\sigma t_\rmP\right) \delta\left(k-\left(k_1-k_2\right)\right) \Theta(t-T)\, \calJ_{\kappa,-\kappa_2}(t,T) \Bigg].
\label{g2k_2_app}
\end{align}
The first and second terms denote parts of the response with wavenumbers $2 k_1$ and $-2 k_2$ respectively. The third term is a cross-term between the $k_1$ and $k_2$ modes and manifests an echo of wavenumber $k_1-k_2$ that occurs slightly after $t=T$. The phase factor in this cross-term becomes independent of $v$, or in other words, phase space bunching occurs at

\begin{align}
t = t_{\rm echo} \approx T + \frac{1}{\nu_\rmc} \ln\left|1+\frac{k_2}{k_1-k_2}\left(1-e^{-\nu_\rmc T}\right)\right|.
\label{t_echo_app}
\end{align}
In the collisionless ($\kappa = k\sigma/\nu_\rmc \to \infty$, $\kappa_2 = k_2\sigma/\nu_\rmc \to \infty$) limit, the phase factor scales as $\sim \exp{\left[-i v\left(\left(k_1-k_2\right) t - k_1 T\right)\right]}$. Accordingly, the $\nu_\rmc T \to 0$ limit of equation~\ref{t_echo_app} implies $t_{\rm echo} \approx \left[1+k_2/\left(k_1-k_2\right)\right]T = k_1 T/\left(k_1-k_2\right)$. In the strong collision or $\nu_\rmc T \to \infty$ limit, $t_{\rm echo} - T \approx \frac{1}{\nu_\rmc}\ln{\left|k_1/\left(k_1-k_2\right)\right|}$. Hence, it becomes increasingly difficult to observe the echo with increasing $\nu_\rmc$.

The echo is a manifestation of the bunching of particles in the phase space. This implies that the macroscopic density perturbation of the $\left(k_1-k_2\right)$ mode (obtained by integrating the cross term in $g_{2\kappa}$ from equation~[\ref{g2k_2_app}] over $u$), given by

\begin{widetext}
\begin{align}
r_{2\kappa} = \frac{\rho_{2k}}{\rho_0} &= - ab \left(k_1\sigma t_\rmP\right) \left(k_2\sigma t_\rmP\right) \delta\left(k-\left(k_1-k_2\right)\right) \Theta(t-T) \nonumber\\
&\times \kappa\left(1-e^{-\nu_\rmc(t-T)}\right) \left[\kappa e^{-\nu_\rmc T}\left(1-e^{-\nu_\rmc (t-T)}\right) - \kappa_2\left(1-e^{-\nu_\rmc T}\right)\right] \nonumber\\
&\times \exp{\left[-\kappa^2\left[\nu_\rmc\left(t-T\right)-\left(1-e^{-\nu_\rmc\left(t-T\right)}\right)\right] -\kappa^2_2\left[\nu_\rmc T-\left(1-e^{-\nu_\rmc T}\right)\right] + \kappa \kappa_2 \left(1-e^{-\nu_\rmc\left(t-T\right)}\right) \left(1-e^{-\nu_\rmc T}\right) \right]}
\label{r2k_app}
\end{align}    
\end{widetext}
peaks around $t = t_{\rm echo}$ (see Fig.~\ref{fig:r2k_vs_t}). 

% The \nocite command causes all entries in a bibliography to be printed out
% whether or not they are actually referenced in the text. This is appropriate
% for the sample file to show the different styles of references, but authors
% most likely will not want to use it.
%\nocite{*}

\bibliography{references_banik}% Produces the bibliography via BibTeX.

\end{document}